\documentstyle[12pt,a4,epsf]{article}

\newcommand{\resection}[1]{\setcounter{equation}{0}\section{#1}}
\newcommand{\appsection}{\addtocounter{section}{1} \setcounter{equation}{0}
                         \section*{Appendix \Alph{section}}}

\def\al{\alpha}

\def\d {\delta}
\def\l {\lambda}

\def\e {\epsilon}
\def\be{\begin{equation}}
\def\ee{\end{equation}}
\def\ba{\begin{array}}
\def\st{\stackrel}
\def\ea{\end{array}}
\def\EQ{\begin{equation}}
\def\EN{\end{equation}}
\def\bea{\begin{eqnarray}}
\def\eea{\end{eqnarray}}
\def\beano{\begin{eqnarray*}}
\def\eeano{\end{eqnarray*}}
\def\to{\rightarrow}
\def\goto{\longrightarrow}

\def\Q{{\cal Q}}

\def\P{{\cal P}}

\def\bout{{}^{out}\langle}
\def\bin{{}^{in}\langle}

\def\kin{\rangle^{in}}
\def\boutz{{}^{out}_{\,\,\,\,\,\,0}\langle}
\def\binz{{}^{in}_{\,\,\,\,0}\langle}
\def\koutz{\rangle^{out}_0}
\def\kinz{\rangle^{in}_0}
\def\lab{\label}
\def\th{\theta}
\begin{document}

\setcounter{page}{0}
\newpage     
\setcounter{page}{0}
\begin{titlepage}
\begin{flushright}
OUTP--96--10S

ISAS/EP/96/23

IC/96/26 

SWAT/95-96/98
\end{flushright}
\vspace{0.5cm}
\begin{center}
{\large {\bf Non-integrable Quantum Field Theories as \\
Perturbations of Certain Integrable Models}}\\
\vspace{1.5cm}
{\bf G. Delfino$^a$, G. Mussardo$^{b,c,d}$ and 
P. Simonetti$^e$ }\\
\vspace{0.5cm}
{\em $^a$ Theoretical Physics, University of Oxford \\
1 Keble Road, Oxford OX1 3NP, UK} \\
\vspace{1mm}
{\em $^b$ International School for Advanced Studies, \\
Via Beirut 3, 34014 Trieste, Italy} \\
\vspace{1mm} 
{\em $^c$ International Center of Theoretical Physics,\\
Strada Costiera 11, 34014 Trieste, Italy} \\
\vspace{1mm}
{\em $^d$ Istituto Nazionale di Fisica Nucleare, \\
Sezione di Trieste, Italy} \\
\vspace{1mm} 
{\em $^e$ Department of Physics, University of Wales Swansea,\\ 
Singleton Park, Swansea SA2 8PP, UK} \\ 
\end{center}
\vspace{3mm}
\begin{abstract}
\noindent
We approach the study of non--integrable models of two--dimensional
quantum field theory as perturbations of the integrable ones.
By exploiting the knowledge of the exact $S$-matrix and Form Factors 
of the integrable field theories we obtain the first order corrections 
to the mass ratios, the vacuum energy density and the $S$-matrix of 
the non-integrable theories. As interesting applications of the formalism, 
we study the scaling region of the Ising model in an external magnetic 
field at $T \sim T_c$ and the scaling region around the minimal model 
$M_{2,7}$. For these models, a remarkable agreement is observed between 
the theoretical predictions and the data extracted by a numerical 
diagonalization of their Hamiltonian. 
\end{abstract}
\vspace{5mm}
\end{titlepage}
\newpage
\section{Introduction}
Two-dimensional integrable and relativistic models have provided a most 
valuable heuristic guide in the analysis of the infinite-dimensional space of 
Quantum Field Theories and the associated statistical models.  
The peculiar property of the dynamics that makes those models solvable 
consists in the existence of an infinite number of integrals of motion. 
This feature manifests itself in several ways \cite{ZZ,Zam}. First of all, it 
leads to very severe selection rules: for instance, scattering processes 
accompanied by production or annihilation events cannot take place in 
such theories and therefore the only allowed scattering processes are 
purely elastic. Secondly, the $n$-particle scattering amplitudes can 
always be factorized into the product of the $n(n-1)/2$ two-particle 
elastic $S$-matrices. The two-particle $S$-matrix $ S_{ab}(s)$ for the 
particles $A_a$ and $A_b$, with mass $m_a$ and $m_b$ respectively, 
has only two elastic branch cut singularities at 
$s = (m_a - m_b)^2$ and $s = (m_a + m_b)^2 $ and fulfills by itself the 
unitarity equation
$
S_{ab}(s) \,S_{ab}^{\dagger}(s) = 1
$ 
in the whole Riemannian surface of the Mandelstam variable $s$. 
Finally, the exact mass spectrum and the complete set of $S$-matrices of 
integrable QFT may be computed according to the so-called 
{\em bootstrap principle} which states that the bound states have to 
be regarded on the same footing as the asymptotic states.  
The drastic simplification of the on-shell properties of the integrable 
field theories has far-reaching consequences since it greatly facilitates 
both the computation of their thermodynamic quantities and the determination 
of their correlation functions (see, for instance [10-20]). 
For instance, the usual lengthy computation of matrix elements of local 
operators which enters the spectral representation of correlation functions 
can be shortened in the case of integrable models to solve a finite number 
of functional and recursive equations. At present, the exact $S$-matrix, the 
exact mass spectrum and the correlation functions of a large number of 
integrable models have been determined\footnote{For a brief list of 
references see [1-20].} and confirmed either by analytic lattice 
computations or by numerical simulations [32-39].
One of the most remarkable examples of statistical model solved by using 
the bootstrap approach is represented by the Ising model in a magnetic field at 
$T=T_c$: the exact $S$-matrix proposed by Zamolodchikov \cite{Zam} has been 
the starting point for the solution of the long-standing problem of determining the 
spin-spin correlation function of this model \cite{DM}.

Despite all the distinguishing qualities of integrable models and 
their successful applications to statistical mechanics, the actual realm of 
particle physics is however that of non-integrable QFT. These theories 
generally present the striking phenomenon of resonances as well as 
multi-production reactions. Many interesting statistical models in the 
vicinity of their fixed points may be described by non-integrable QFT and, to make 
any progress in the computation of their thermodynamical quantities (or, 
simply to understand better their qualitative features), it would be highly 
desirable to develop an appropriate formalism to deal with the lack 
of integrability. This task is notoriously difficult because 
the rich physical scenario of the multichannel systems, represented by 
the non--integrable field theories, is usually accompanied by great 
mathematical complexities. In fact, once one has given up the integrability 
condition, the infinite number of thresholds associated to the production 
processes greatly influences the analytic structure of the scattering 
amplitudes, inducing a rich pattern of branch cut singularities in addition 
to the pole structure generally associated to the bound or resonance states 
(Figure 1). 

The mathematical difficulties of the problem are well known and cannot 
be easily circumvented, as illustrated by the following line of reasoning.  
Let $T_1 < T_2  \ldots < T_n \ldots $ be the infinite sequence of 
energy thresholds of the scattering amplitudes entering the $S$-matrix.  
In each interval $T_n \leq s < T_{n+1}$, 
the physical $S$-matrix is given by an $n\times n$ unitary matrix
$S_{ij}^{(n)}(s)$ ($i,j=1,\ldots,n$), where here the suffixes 
$i$ and $j$ are multi-particle collective indices that generally label 
the different channels of the scattering process. At each 
threshold, a new channel opens up and correspondingly, more matrix elements 
have to be introduced. Due to the nested structure of the $S$-matrix 
as a function of $s$, i.e. 
$S^{(1)} \rightarrow S^{(2)} \ldots \rightarrow S^{(n)} $,
one may be inclined to approach the difficult problem of computing the 
infinite number of scattering amplitudes by applying a recursive procedure. 
Namely, assuming that all the $n^2$ matrix elements $S^{(n)}_{ij}(s)$ 
have already been determined as analytic functions of $s$ in the interval 
$T_n \leq s < T_{n+1}$, one might try to analytically continue them into 
the next interval $T_{n+1} \leq  s < T_{n+2}$ and then, as consequence,  
compute the new unknown scattering amplitudes entering the larger 
matrix $S^{(n+1)}_{ij}$ by using the unitarity equations relative to the 
new range of energies. There is, of course, an important flaw in the above 
argument:  the recursive procedure is in fact based on the assumption that 
the matrix elements $S^{(n)}_{ij}(s)$ can be prolongated through the different thresholds 
without being halted by the presence of singularities, an assumption which 
is obviously false. Hence, the complexity of non-integrable QFT consists 
in the fact that in any energy interval one must take {\em simultaneously} 
into account {\em all} the inelastic thresholds of the theory. 
This feature is explicitly shown by a simple example discussed in 
Appendix A. For a discussion on the inter-related aspects 
of analyticity and integrability, see also ref.\,\cite{chiral}. 

The problem of analysing non-integrable field theories (or at least a class 
thereof) is not however as impracticable as it may seem at first sight. 
A possible approach to their study is suggested by the observation that 
although integrable and non-integrable QFT possess quite different properties 
as far as scattering amplitudes and mass spectrum are concerned, their 
ultraviolet limit may be described by the {\em same} Conformal Field Theory 
(CFT). From this point of view, an integrable field theory is nothing more 
than a {\em particular deformation} of a Conformal Field Theory whose 
euclidean action may be written as  
\EQ
{\cal A}^{i}_{\rm int} \,=\, {\cal A}_{\rm CFT} + 
g \int\, d^2x \, {\Phi_i}(x) \,\, ,
\label{deformation1}
\EN
where ${\Phi_i}(x)$ is one of the relevant operators which mantains 
the original integrability of the conformal model\footnote{Zamolodchikov 
has outlined some criteria on the integrability of the deformations of CFT.
For a generic minimal model, for instance, integrable QFT are obtained 
as deformations by the relevant operators $\Phi_{1,3}$, $\Phi_{1,2}$ 
and $\Phi_{2,1}$ \cite{Zam}.}. Starting now from 
a given conformal model, there are at least two ways of defining a 
non-integrable field theory: (a) the first possibility is to simply add 
a perturbation induced by a non-integrable operator to the CFT action ; (b) 
an alternative possibility is to deform the CFT action by means of a linear 
combination of several operators $\Phi_i$, in which at least one of these 
or all of them are individually integrable\footnote{In the case when 
all the operators $\Phi_i$ are individually integrable but their 
simultaneous presence renders the field theory non-integrable, 
it can be argued that this is due to different null-vector structure 
of the fields $\Phi_i$ \cite{GM} or it can also be inferred by numerical 
studies of the spectrum, as discussed in the next 
sections.}. For the corresponding action in the last case we have 
\EQ
{\cal A} = {\cal A}_{\rm CFT} + \sum_{j} g_j \int \,d^2 x\, \Phi_{j}(x)
\,\,\,.
\label{multiple}
\EN  

Apart from numerical investigations \cite{Marcio} or perhaps standard 
analyses 
based on conformal perturbation methods, non-integrable QFT of type (a) are 
presently difficult to study and they will not be considered in this paper. 
On the contrary, non-integrable QFT of type (b) will be shown in this 
paper to be suitable of a purpose of theoretical analysis. This is 
particularly important because they provide in many respects 
the most interesting realization of non-integrable models. Let us 
illustrate the key observation that permits their analytic approach. 

In order to study the non-integrable QFT associated to the euclidean 
action (\ref{multiple}), it is convenient to regard the latter 
as the action relative to {\em a deformation of an integrable QFT} rather 
than a multiple deformation of CFT. This consists in grouping differently 
the terms in (\ref{multiple}) and rewriting it as 
\EQ
{\cal A} = {\cal A}^{i}_{\rm int}
+ \sum_{j\neq i} g_j \int \,d^2x \,\Phi_j(x) \,\,\, .
\label{group}
\EN
There are several advantages to adopting this point of view. 
\begin{itemize}
\item Firstly, going to the Minkowski space, we can start our analysis of 
non-integrable QFT by using the particle basis of the integrable model 
associated to ${\cal A}_{\rm int}^{i}$. Although their mass spectrum will not 
generally coincide, the particle basis of the latter model is surely more 
suitable than the conformal basis for capturing the large distance properties 
of the non-integrable theory\footnote{This is generally the case for massive 
field theories. For massless non-integrable field theories the analysis  
may be more subtle and will not be pursued here.}. 
\item Secondly, the integrable QFT with which we start our 
analysis are solvable theories, although not necessarly free. This means 
that we know how to compute the exact matrix elements of all their local 
operators, in particular those entering the action (\ref{group}). 
Hence, we can develop in this case a perturbative approach to 
non-integrable QFT based on the exact Form Factors of the integrable ones. 
As we will discuss in this paper, this 
approach seems generally more powerful and efficient than the one based on 
usual conformal perturbation theory. In fact, a large set of quantitative 
information --concerning the mass corrections, the renormalization of the 
vacuum energy density and the $S$-matrix of the non-integrable models-- can 
already be obtained at the lowest order in the coupling constant $g_j$ with 
a high level of reliability\footnote{The formulas which we will 
derive in Section 3 can be regarded as ``Born approximation formulas" for 
a non-integrable QFT.}. 
\item Finally, in the case of multiple deformation of CFT by separately 
integrable operators, we can choose {\em any } of them for defining our 
initial solvable QFT. Obviously, each choice leads to a selection of particle basis, 
bound state structure and dynamics, on which we can construct our 
perturbative approach. However, since the actual dynamics of non-integrable 
QFT must be independent of the solvable model we start with, this implies 
that perturbative expansions based on different Form Factors and spectra 
must be related each other by a set of mathematical identities. 
\end{itemize}

The most natural interpretation of eq.\,(\ref{multiple}) when all 
$\Phi_i(x)$ are individually integrable operators, is then that of an 
interpolating action between exactly solvable models, the interpolation 
being realized by varying the relative values of the coupling constants 
$g_i$ (Figure 2). A remarkable example of such interpolating theory is 
provided by the Ising model away from the critical point, i.e. at 
$T \neq T_c$ and in a presence of a magnetic field $h$. In the pure 
magnetic direction, the spectrum of this model consists of 
eight stable particles, five of which above threshold \cite{Zam}. In the pure 
temperature direction, on the contrary, the spectrum presents only one massive 
state, which can be interpreted either as a kink (in the low-temperature 
phase) or as a particle (in the high-temperature phase) [22-26]. 
The interpolation between the Hilbert spaces constructed on these rather 
different bases and the rich physical scenario associated to it will be 
analysed in Section 5.  

At this stage of the discussion, the careful reader may have noticed that 
we could have directly introduced the non-integrable QFT by means of the 
action (\ref{group}) without further specifying how the integrable 
action ${\cal A}_{\rm int}^{i}$ has been actually obtained. This is indeed 
true and in fact, all the formalism set up in this paper to study 
non-integrable models basically relies on the properties of relativistic 
integrable models as defined, for instance, by the framework of the analytic 
theory of the $S$-matrix. Although the perturbed CFT approach we have used in 
this section may then appear only as a convenient way to introduce and define 
non-integrable QFT, however the eventual knowledge of the underlying 
conformal model actually adds an important piece of information on the 
operator content and the
ultraviolet properties of the theory. In fact, by the knowledge of the CFT 
ruling the short-distance Operator Product Expansion of the theory, one can 
severerly constrain its high-energy behaviour and obtain, among other things, 
a bound on the cross-sections of non-integrable QFT. This aspect as well as 
others of non-integrable models will be discussed in a separate 
publication \cite{preparation1}.  

The paper is organized as follows. In Section 2 we define the perturbative 
scheme based on the so-called intermediate state representation and the 
Form Factors of integrable theories. We also discuss the necessity of 
introducing some counterterms in the perturbative series in order to 
have a consistent formulation of the theory. In Section 3 we specify our 
analysis to the case of two-dimensional integrable QFT and derive the main 
formulas which rule the variation of the mass ratios, the vacuum energy 
density and the $S$-matrix. Section 4 is devoted to the simplest model which 
can be discussed by means of these techniques, namely the minimal model 
${\cal M}_{2,7}$ perturbed by the operators $\Phi_{1,2}$ and $\Phi_{1,3}$. 
Another example of non-integrable quantum field theory is given by the Ising 
model away from criticality, at a generic point in the bidimensional phase 
diagram spanned by the magnetic field $h$ and the reduced temperature 
$(T - T_c)/T_c$. In view of its important role in statistical mechanics, 
the Ising model away from criticality will be discussed in some detail 
in Section 5. In particular, we will compute the first order corrections 
to the spectrum of the theory when a small thermal perturbation is added to
the pure magnetic case. Our conclusions are then reported in Section 6. 
Three appendices complete the paper: the first one shows how the inelastic 
thresholds enter the elastic scattering amplitude of a simple example of 
quantum mechanics; the second appendix discusses the Faddeev-Zamolodchikov 
algebra and the disconnected terms present in the matrix elements of local 
operators; the last one gathers the relevant Form Factor expressions for 
the $\Phi_{1,3}$ deformation of the $M_{(2,7)}$ model.

\section{The Intermediate State Representation}

Let us consider a quantum field theory in the Minkowski space, 
defined by the action
\EQ
{\cal A} = {\cal A}_0 + {\cal A}_I = {\cal A}_0 - \l \int d^2x \,\Psi(x)\,\,,
\lab{action}
\EN
where ${\cal A}_0$ denotes here the minkowskian action of the unperturbed theory
and $\Psi$ one of its operators. We suppose that the QFT associated to the 
action ${\cal A}_0$ is exactly solvable (although not necessarily free), i.e. 
we assume that the spectrum of particles, their scattering amplitudes and
the matrix elements of the operators of the theory (and in particular those 
of $\Psi$) are all known. For the sake of simplicity, we consider in this
section the case of isospectral perturbations of a solvable theory.
This means that the spectrum of the theory described by the total action 
${\cal A}$ will be made of the same number of particles of the
unperturbed one; in other words, the new interaction ${\cal A}_I$ 
only effects the values of the 
masses of the physical particles but not their stability 
properties\footnote{We will comment on the more general case 
in Section 5 when we will consider the Ising model.}. 

Let us now describe the properties of the theory associated to the action 
${\cal A}$. Under the hypothesys that the interaction term is turned off at 
$t \to \pm \infty$, it is possible to adopt the formalism of the asymptotic 
``{\em in}" and ``{\em out}" states. We are interested in computing 
the scattering amplitude
\be
S \{ q_1,\ldots, q_n\to q'_1,\ldots, q'_m \} = \bout 
q'_1,\ldots,q'_m | q_1,\ldots,q_n\kin = \bin q'_1,\ldots,q'_m | S | 
q_1,\ldots,q_n\kin \,\, ,
\lab{smatr}
\ee
where ${q_i}$ and ${q'_j}$ label the momenta of the {\em in}-going and
{\em out}-going set of 
particles. Since in the remote past $t\to -\infty$ the interaction is not 
present yet, the asymptotic {\em in} states coincide with the 
unperturbed ones
\be
|q_1,\ldots,q_n\kin = |q_1,\ldots,q_n \kinz \,\,\,.
\ee
As usual, the scattering operator $S$ appearing in (\ref{smatr}) 
can be obtained as the limit 
\EQ
S=\lim_{t \to +\infty}U(t,-t)\,\, 
\EN
of the time evolution operator $U(t,t_0)$, which is the solution of the
equation
\EQ
i\frac{d}{dt}U(t,t_0) \,=\,H \,U(t,t_0)\,\,,\hspace{1cm}U(t_0,t_0)=1\,\,,
\lab{qm}
\EN
where $H=H_0+H_I$ denotes the Hamiltonian associated to the theory
(\ref{action}). Following the standard quantum mechanical procedure
(see, for instance \cite{messiah}), the operator $U$ can be factorised
as $U=U_0U_I$, where $U_0$ and $U_I$ are solutions of eq.\,(\ref{qm})
with $H$ replaced by $H_0$ and $\tilde{H}_I(t)=U_0^{-1} H_I U_0$, 
respectively. Then, we can write the scattering operator of the theory 
(\ref{action}) as $S = S_0 S_I$, where $S_0=\lim_{t\to +\infty}U_0(t,-t)$ 
is the unperturbed and exactly known scattering matrix
\be
S_0 \{ q_1,\ldots,q_n \to q'_1,\ldots,q'_m\} = 
\boutz q'_1,\ldots,q'_m| q_1,\ldots,q_n\kinz = 
\binz q'_1,\ldots,q'_m | S_0 | q_1,\ldots,q_n \kinz \,\, ,
\lab{s0}
\ee
while $S_I$ has the usual formal representation
\be
S_I = \lim_{t \to +\infty} U_I(t,-t) = T\exp \left( i A_I[\Psi] \right)\,\,\,.
\ee
The scattering amplitude is therefore given by 
\bea
&& \bout q'_1,\ldots,q'_m | q_1,\ldots,q_n \kin  =  
\boutz q'_1,\ldots,q'_m | T \exp \left( - i \l \int d^2x \Psi(x)  \right)
| q_1,\ldots,q_n \kinz = \nonumber \\
& & \,\,\,= \sum_{k=0}^{+\infty} \frac{(- i \l)^k}{k!} \int   d^2x_1 \ldots d^2x_k 
\boutz q'_1,\ldots,q'_m|
T\left(\Psi (x_1) \ldots \Psi(x_k)\right) | q_1,\ldots q_n\kinz\,\,,
\lab{amplitudes}
\eea
where (\ref{s0}) has been used in order to absorb the factor $S_0$.

In ordinary lagrangian perturbation theory based on free theories, the
computation of scattering amplitudes would now proceed through the use
of creation and annihilation operators and Wick's theorem, finally
leading to the diagrammatic expansion characteristic of Feynman
covariant perturbation theory. Although this approach cannot be 
generally followed here (because we might not know, in general, a local 
lagrangian formulation of the theory associated to ${\cal A}_0$), 
the knowledge of the exact solution of its dynamics, in the form 
specified at the beginning of this section, naturally suggests to 
compute the scattering amplitudes (\ref{amplitudes}) within a 
framework quite analogous to that of ordinary time-dependent 
perturbation theory in quantum mechanics.
In other words, we initially insert between the operators $\Psi(x_l)$ and
$\Psi(x_{l+1})$ ($l=1,\ldots,k-1$) in the second line of
(\ref{amplitudes}) a sum over a complete set of asymptotic states of
the unperturbed theory
\be
\sum_n | n \kinz \ \binz n | = 1 = \sum_n | n \koutz \ \boutz n |\,\,,
\ee
with $|n\rangle$ denoting an asymptotic state containing $n$ on--shell
particles. Then, the integrations on the space coordinates in 
(\ref{amplitudes}) are immediately performed and lead to delta functions 
constraining the total momentum of the intermediate states to coincide 
with that of the initial and final states. In doing the integrations on the 
time variables, the time ordering prescription gives rise in this case 
to the appearence of energy denominators so that one ends up with the 
following expression
\bea
&& \bout q'_1,\ldots,q'_m | q_1,\ldots,q_n \kin = 
\boutz q'_1,\ldots,q'_m | q_1,\ldots,q_n \kinz \,\,
+ \nonumber \\
&& \,\,\,+ (2 \pi)^2 \d^{(2)}\left(\sum_{j=1}^m q'_j-\sum_{j=1}^n q_j\right) 
\, 
\left\{ -i\l \,\,\,\boutz q'_1,\ldots,q'_m |\Psi(0)| q_1,\ldots,q_n \kinz 
+ \right. \label{pertform} \\
&&
+ \frac{1}{2\pi i} 
\sum_{k=2}^{+\infty} (2\pi\l)^k \sum_{n_1}\ldots\sum_{n_{k-1}} \left[  
\frac{\d(Q - P_1)\ldots\d(Q - P_{k-1})}
{(E - E_1 + i \e) \ldots (E - E_{k-1} + i \e )} \times 
\right. \nonumber \\ 
& & \,\,\,\times \left.\left.
\boutz q'_1,\ldots,q'_m |\Psi (0)|n_1\rangle^{}_0 \ldots 
{}^{}_0 \langle n_{k-1}|\Psi(0)| q_1,\ldots,q_n \kinz \right]\right\}  
\nonumber \,\, ,
\eea
where $E$ and $E_i$ ($Q$ and $P_i$) denote the total energy (momentum)
of the initial state and of the $i$--th intermediate state, respectively, and
each intermediate sum can be equivalently taken either on the basis of 
the {\em in} states or on that of {\em out} states.
Since the matrix elements between asymptotic states of the perturbing 
operator $\Psi(x)$ are supposed to be known, the scattering amplitudes
in the quantum field theory (\ref{action}) are in principle computable
quantities, order by order in the coupling constant $\lambda$. The above 
expansion over intermediate states must be contrasted with the usual formalism
of covariant perturbation theory in which both energy and momentum are
conserved in the internal lines of Feynman diagrams where, however, such 
lines correspond to off-shell particles.

As it is, the above formula (\ref{pertform}) is not completely correct though. 
In fact, the new interaction changes both the vacuum energy density and the 
mass of the particles. Hence, we have to refine the action ${\cal A}_I$ by 
introducing some counterterms to take properly into account the correct 
normalisation of the states. We require the validity of the following 
normalisation conditions for any value of the coupling constant: 
the normalisation of the vacuum state
\EQ
\langle 0 | 0\rangle =  
{}_0\langle 0 | 0\rangle_0 = 1 \,\,,
\label{vacnorm}
\EN
and the normalisation of the one-particle states
\EQ
\bout q' | q \kin = 
\boutz q' | q \kinz = 2 \pi E \, \d (q'^1 - q^1)\,\,\,.
\label{onepnorm}
\EN
The two above conditions should be enforced order by order in perturbation 
theory when using (\ref{pertform}) to compute the vacuum to vacuum and the 
one-particle amplitudes. 

Equation (\ref{vacnorm}) leads to subtract a constant term 
$\delta{\cal E}_{vac}(\lambda)$ from the interaction density. This extra 
term obviously measures the variation of the vacuum energy density under
the effect of the perturbation. This variation is usually ignored in
lagrangian perturbation theory where the prescription of disregarding
the disconnected vacuum bubble diagrams is adopted. We keep track of
it here because it is a measurable quantity for the class of models we are
considering in this paper.

The correct one-particle normalisation may be obtained by introducing 
a `mass' term operator in the interaction density. This operator, denoted 
here by $O^{(2)}(x)$, can be defined in terms of its (unperturbed) 
Form Factors, given by 
\be
F_n^{O^{(2)}} = {}_0\langle 0|O^{(2)}(0)|q_1,\ldots,q_n \kinz=\d_{n,2} \,\,.
\label{masscount}
\ee
With this definition, the coefficient in front of the operator $O^{(2)}(x)$ 
in the interaction density has to be determined by imposing 
eq.\,(\ref{onepnorm}) order by order in the coupling $\l$ and plays the 
role of a mass counterterm $\delta m^2(\lambda)$.

In summary, the correct formula for the scattering amplitude is given by  
\bea
& &\bout q'_1,\ldots,q'_m | q_1,\ldots,q_n \kin = 
\boutz q'_1,\ldots,q'_m | q_1,\ldots,q_n \kinz \,\,\, + \nonumber \\
&& -i\,(2 \pi)^2 \d^{(2)}\left(\sum_{j=1}^mq'_j-\sum_{j=1}^n q_j\right)\times
\nonumber \\ 
& & \times\left\{ \,\,\,\boutz q'_1,\ldots,q'_m | 
\left( \l \Psi (0) - \frac{1}{2}\d m^2 O^{(2)}(0) -
\d{\cal E}_{\rm vac}\right)| q_1,\ldots,q_n \kinz + \right.
\label{start} \\  
& & +\sum_{n_1} \frac{2\pi \,\d(Q - P_1)}{(E - E_1 + i \e)} 
\boutz q'_1,\ldots,q'_m | \left( \l \Psi (0) - 
\frac{1}{2}\d m^2 O^{(2)}(0) -
\d{\cal E}_{\rm vac} \right)
|n_1 \rangle^{}_0 \times \nonumber \\
&& \left.{}^{}_0 \langle n_1|
\left( \l \Psi (0) - \frac{1}{2}\d m^2 O^{(2)}(0) -
\d{\cal E}_{\rm vac}\right)|q_1,\ldots,q_n \kinz \,\,+ 
\ldots {}_{}^{}\right\} \nonumber \,\,\,.
\eea

Let us remark that the above expansion appears as the most physical one since 
it deals with the true physical degrees of freedom of the problem. 
However, as usual in quantum field theory appearences of divergent 
contributions are expected when the above formula is applied beyond 
the first perturbative order. A general discussion of such divergences and of the 
renormalisation procedure which must be adopted to deal with the infinities, 
seems to be an interesting open problem in the unconventional setting we are 
considering. In fact, it must be taken into account that for the perturbations 
of the interacting theories we are interested in, the perturbing operator 
$\Psi(x)$ has in general non-vanishing matrix elements on {\em all} the 
asymptotic states. This means that a resummation of an infinite number of terms
is required at any perturbative order beyond the first. The general 
aspects of this problem will not however be further investigated in this paper. 
Rather, we will concentrate our attention here on the first order approximation 
in order to study the effects induced on an integrable model of 
two--dimensional quantum field theory by a small perturbation which 
breaks the integrability.

\resection{Perturbing Two-dimensional Integrable Models}

In this section we apply the perturbative scheme exposed in the previous 
section to the case in which the term ${\cal A}_0$ in the euclidean version 
of the action (\ref{action}) corresponds to an {\em integrable} deformation 
of a two-dimensional conformal field theory through a relevant operator 
$\Phi(x)$, i.e. 
\EQ
{\cal A}_0 = {\cal A}_{CFT} + g\int d^2x\,\Phi(x)\,\,\,.
\lab{int}
\EN
Before proceeding in our analysis, a brief review of the main properties 
of the integrable relativistic field theories on which we base our 
sequent considerations seems in order. 

\subsection{Basic Features of Integrable Theories}

Integrable theories are characterized by the existence of an infinite 
number of integrals of motion. This circumstance gives the possibility 
of solving them along completely non-perturbative methods. The first 
consequence of the infinite number of conserved currents is that the 
scattering processes in the integrable theory are completely elastic, i.e. 
the final state contains the same number of particles with the same momenta 
of the initial one: the only results of the scattering processes are 
then a possible exchange of the ``color" quantum numbers between particles 
in the same mass multiplet together with the elastic phase shifts 
associated (semi-classically) to the time delays of the wave packets 
\cite{ZZ}. Moreover, in these cases the S-matrix is completely factorised 
into the product of the two particle scattering amplitudes defined by 
\EQ
{}^{out}_{\,\,\,\,\,\,0}\langle c(\th_3)d(\th_4)|a(\th_1)b(\th_2)
\rangle_0^{in}=
(2\pi)^2\delta(\th_1-\th_3)\delta(\th_2-\th_4)S_{ab}^{cd}(\th_1-\th_2)\,\, ,
\EN
where the rapidity variable $\th_i$ parameterises the on-mass shell 
energy and momentum of the particles as
$(p^0_i,p^1_i) = (m_i\cosh\th_i,m_i\sinh\th_i)$. Since the Mandelstam
variable $s$, given by 
\EQ
s_{ab}(\th_1-\th_2)=(p_a(\th_1)+p_b(\th_2))^2=
m_a^2+m_b^2+2m_am_b\cosh(\th_1-\th_2)\,\,,
\lab{mand}
\EN
is the only independent relativistic invariant in the process, the
scattering amplitude depends on the rapidity difference only. Once
combined with the general principles of analyticity, unitarity and
crossing symmetry, the elasticity and factorisation properties usually
lead to the exact determination of the $S$-matrix in integrable models
(see for instance [1-9]). 

The second important characteristic of the two-dimensional relativistic 
integrable models is that, in addition to the $S$-matrix, we can 
also determine exactly the matrix elements of the local operators 
${\cal O}(x)$ of the theory on the asymptotic states, i.e. 
\EQ
{}_{b_1\ldots b_m}F^{\cal O}_{a_1\ldots
a_n}(\th'_1,\dots,\th'_m|\th_1,\ldots,\th_n) \equiv 
{}^{out}_{\,\,\,\,\,\,0}\langle b_1(\th'_1)\ldots b_m(\th'_m)|{\cal O}(0)|
a_1(\th_1)\ldots a_n(\th_n)\rangle_0^{in}\,\,\,.
\lab{matrixelements}
\EN
We can always restrict our attention to those matrix elements with no 
particle on the left side, the so-called Form Factors (FF) 
\EQ
F^{\cal O}_{a_1\ldots a_n}(\th_1,\ldots,\th_n)={}_0\langle 0|{\cal O}(0)
|a_1(\th_1)\ldots,a_n(\th_n)\rangle^{in}_0 \,\,\,.
\label{FormFactor}
\EN
In fact, the generic matrix element (\ref{matrixelements}) can be obtained 
in terms of the Form Factors through the analytic continuation
\bea
&& {}_{b_1\ldots b_m}F^{\cal O}_{a_1\ldots
a_n}(\th'_1,\dots,\th'_m|\th_1,\ldots,\th_n) =  
F^{\cal O}_{\bar{b}_1\ldots\bar{b}_ma_1\ldots
a_n}(\th'_1+i\pi,\ldots,\th'_m+i\pi,\th_1,\ldots,\th_n) \nonumber\\
&& \,\,\,\,\,\,\,\,\,\,\, + \,{\rm disconnected \,\, parts}\,\,,
\lab{continuation}
\eea
where the bar denotes the charge-conjugated particles. The ``disconnected 
parts'' in the right hand side of (\ref{continuation}) obviously appear 
when some of the primed rapidities coincide with the unprimed ones 
(this point is discussed in more detail in appendix B). 

As well known, the computation of the Form Factors can be performed once the 
exact $S$--matrix and the bound state structure of the theory are 
known. In fact, the monodromy properties 
of the FF are constrained by the so--called Watson equations, given by 
\begin{eqnarray} 
F^{\cal O}_{a_1\ldots a_ia_{i+1}\ldots a_n}(\theta_1,\ldots,
\th_i,\th_{i+1},\ldots,\th_n)\!\!\!&=&\!\!\!   \nonumber \\
  & &\!\!\!\!\!\!\!\!\!\!\!\!\!\!\!\!\!\!\!\!\!
  \!\!\!\!\!\!\!\!\!\!\!\!\!\!\!\!\!\!\!\!\!
   = S_{a_ia_{i+1}}^{b_ib_{i+1}}(\th_i-\th_{i+1})
\:F^{\cal O}_{a_1\ldots b_{i+1}b_i\ldots a_n}(\theta_1,\ldots,
\th_{i+1},\th_{i},\ldots,\th_n)\, ,\nonumber
\end{eqnarray}
\EQ \label{sf}
F^{\cal O}_{a_1a_2\ldots a_n}(\theta_1+2\pi i,\th_2,\ldots,\th_n)= 
F^{\cal O}_{a_2\ldots a_na_1}(\th_2, \ldots, \th_n, \theta_1)\, \, ,
\EN
whereas their analytic structure is closely related to the underlying 
pattern of singularities induced by the (multi)-scattering processes 
\cite{DM,KW,Smirnov} and is ruled by a set of recursive equations. The 
simplest of these equations\footnote{For simplicity we quote the formula 
corresponding to the case of diagonal scattering.} are obtained by the 
residue of the FF at the kinematical poles of particle--antiparticle 
singularities of relative rapidity $\theta = i \pi$ \cite{Smirnov} 
\be 
\label{kinpole}
-i\lim_{\tilde{\th}\rightarrow\th}F^{\cal O}_{\bar{a}aa_1\ldots a_n}
(\tilde{\th}+ i\pi,\th,\th_1,\ldots,\th_n)=\left(1-\prod_1^n S_{aa_i}(\th-\th_i)\right)
F^{\cal O}_{a_1\ldots a_n}(\th_1,\ldots,\th_n)\,\,\,.
\ee
Additional recursive equations are induced both by the bound-state poles and 
higher-order poles due to multi-scattering virtual processes 
\cite{DM,KW,Smirnov}. Finally, the asymptotic behaviour of the FF for 
very large values of the rapidities can be controlled according 
to a simple criterion proposed in \cite{DM}: denoted by 
$x_{\cal O}$ the scaling dimension of the operator ${\cal O}(x)$ 
and by $y_{\cal O}$ the real quantity defined by 
\[
\lim_{\mid \th_i\mid \rightarrow \infty} F^{\cal O}
_{a_1\ldots a_n}(\th_1,\ldots,\th_n)\,\sim e^{y_{\cal O}\mid\theta_i\mid}
\]
we have  
\EQ 
\label{boundff}
y_{\cal O} \leq \frac{x_{\cal O}}{2} \,\,\,.
\EN

The equations and the constraints which we have briefly illustrated above 
prove in general sufficient to determine the FF of the local operators of 
the integrable theories (see, for instance [11-20]).

\subsection{First Order Perturbation Theory}

Let us now deform the integrable action (\ref{int}) by adding to it the 
relevant operator $\Psi(x)$. Both the operators $\Phi(x)$ and 
$\Psi(x)$ are assumed to be scalar and relevant, with their scaling 
dimensions denoted by $x_\Phi$ and $x_\Psi$, respectively. 
The theory (\ref{action}) depends in this case on the two dimensionful 
couplings constants\footnote{For the sake of simplicity of notation, 
we assume that no other interaction term is generated by renormalisation 
effects (this is the case of the two models which we will explicitely 
discuss in this paper), although the first order results of this section 
do not actually depend on this assumption.} $g$ and $\lambda$. Since 
$g\sim M^{2-x_\Phi}$ and $\l\sim M^{2-x_\Psi}$, where $M$ is a mass scale, 
we can decide to use $g$ as dimensionful parameter of the theory and 
the dimensionless combination
\EQ
\chi \equiv \l \,g^{-\frac{2 - x_\Psi}{2 - x_\Phi}} 
\EN
as a label of the different Renormalization Group trajectories 
which originate from the fixed point at $g = \lambda = 0$ (see Fig.\,2).  
For example, if $N(\chi)$ denotes the number of stable particles in the 
spectrum of the theory, their masses may be expressed as
\EQ
m_a(g,\chi)={\cal
C}_a(\chi)\,g^{\frac{1}{2-x_\Phi}}\,\,,\hspace{1cm}a=1,2,\ldots,N(\chi)\,\, ,
\lab{ma}
\EN
where ${\cal C}_a(\chi)$ is an amplitude which characterises the whole 
trajectory. Similarly, the vacuum energy density can be written as
\EQ
{\cal E}_{\rm vac}(g,\chi) = {\cal D}(\chi)\, g^{\frac{2}{2-x_\Phi}}\,\,\,.
\lab{evac}
\EN
Dimensionless quantities, as for instance mass ratios, only depend on 
$\chi$ and therefore they do not vary along the trajectories of the 
Renormalization Group. 

Once the new interaction $\l \int d^2x \,\Psi(x)$  is switched on in the 
action, the integrability characterising the unperturbed theory is generally 
lost and the S-matrix becomes extremely more complicated. Inelastic processes 
of particle production are no longer forbidden and the analytic structure 
of the scattering amplitudes will present additional cuts due to the 
higher thresholds. In particular, their expression is no longer factorized 
into the sequence of two-body scattering amplitudes and, even in elastic 
channels, the only surviving restriction on the final momenta comes from 
energy-momentum conservation. 

The knowledge of the matrix elements (\ref{matrixelements}) 
of the perturbing field $\Psi(x)$ ensures the possibility to compute 
perturbatively the amplitudes of the inelastic processes as well as the 
corrections to the elastic ones. To the first order in $\l$ and with an 
obvious extension of the notation, equation (\ref{start}) reads
\bea
&& {}^{out}\langle b_1(q_1^1)\ldots b_m(q_m^1)|a_1(p_1^1)\ldots
a_n(p_n^1)\rangle^{in}\simeq \\
&& \hspace{3mm} \simeq 
\delta_{mn}\,{}^{out}_{\,\,\,\,\,\,0}\langle
b_1(q_1^1)\ldots b_n(q_n^1)|a_1(p_1^1)\ldots
a_n(p^1_n)\rangle_0^{in} + \nonumber
\lab{firstorder}\\
&& \hspace{3mm} 
-i\delta^2\left(\sum_{k=1}^n p_k^\mu-\sum_{k=1}^m q_k^\mu\right)
{}^{out}_{\,\,\,\,\,\,0}\langle
b_1(q_1^1)\ldots b_n(q_m^1)|\l\left(\Psi(0) + \right.\\
&& \,\,\,\hspace{3mm} - 
\left.\frac{1}{2}\sum_{a,b=1}^N\delta
M_{ab}^2 \,O_{ab}^{(2)}(0)-\delta{\cal E}_{\rm vac} \right)|
a_1(p_1^1)\ldots a_n(p^1_n)\rangle_0^{in}\,\,\,. \nonumber
\eea
The ``mass operator'' $O_{ab}^{(2)}(x)$ is defined assigning its form
factors, which with an obvious generalization of eq.(\ref{masscount}), 
are given by 
\EQ
F^{O_{ab}^{(2)}}_{a_1\ldots
a_n}(\th_1,\ldots,\th_n)=\delta_{n2}\delta_{aa_1}\delta_{ba_2}\,\,\,.
\EN
The first order corrections to the masses of the particles and to the
vacuum energy density are then obtained imposing the conditions 
(\ref{onepnorm}) and (\ref{vacnorm}). As a result, 
we have 
\EQ
\delta M_{\bar{b}a}^2\, \simeq\, 2\l \,F^\Psi_{\bar{b}a}(i\pi,0)
\, \delta_{m_am_b}\,\,,
\lab{dm}
\EN
\EQ
\delta {\cal E}_{\rm vac}\simeq\l\,\,\left[
{}_0\langle 0|\Psi|0\rangle_0\right]\,\,\,.
\lab{de}
\EN

Care must be however taken when using the rapidity parameterisation in
eq.\,(\ref{firstorder}). Let's illustrate this point by considering the
first order correction to some elastic process $ab\to cd$. In the
unperturbed theory, this process is characterised by the scattering
amplitude $S_{ab}^{cd}(\th)$, where $\th=\th_1-\th_2$ denotes the
rapidity difference of the colliding particles. In two dimensions, the
momenta of the particles in a two-body elastic collision are
individually conserved even in absence of integrability, so that the
general elastic amplitude $S_{ab}^{cd}(\th,\chi)$ can be introduced through
the relation
\EQ
{}^{out}\langle c(\th_1)d(\th_2)|a(\th_3)b(\th_4)\rangle^{in}=
(2\pi)^2\delta(\th_1-\th_3)\delta(\th_2-\th_4)\,S_{ab}^{cd}
(\th_1-\th_2,\chi)\,\,\,.
\EN
Notice, however, that away from the integrable direction (i.e. $\chi=0$), 
the scattering amplitude $S_{ab}^{cd}(\th,\chi)$ is no longer a meromorphic 
function of $\th$ since the opening of inelastic channels induces an 
additional 
analytic structure. When we compute the correction to $S_{ab}^{cd}(\th,\chi)$
around $\chi=0$, we must take into account that, since the total energy of 
the colliding system is fixed, the variation in the masses given by 
eq.\,(\ref{dm}) induces a corresponding change in the rapidity difference 
expressed by 
\EQ
\delta\th\simeq-\,\frac{m_a\delta m_a+m_b\delta m_b+ 
(m_b\delta m_a+m_a\delta m_b)\cosh\th}{m_am_b\sinh\th}\,\,\,.
\lab{dth}
\EN
Then the correction to the amplitude can be decomposed as
\EQ
\delta S_{ab}^{cd}(\th,\chi)=\frac{\partial
S_{ab}^{cd}(\th)}{\partial\th}\,\delta\th+\left.\frac{\partial
S_{ab}^{cd}(\th,\chi)}{\partial\chi}\right|_{\chi=0}\,\delta\chi\,\,\,.
\lab{total}
\EN
The first order result for this quantity is obtained by using formula
(\ref{firstorder}). Taking into account the cancellation occurring
between the disconnected parts in eq.\,(\ref{continuation}) and the
contributions of the counterterms, one finally obtains 
\EQ
\delta S_{ab}^{cd}(\th,\chi)\simeq
-i\l\,\frac{F^\Psi_{\bar{c}\bar{d}ab}(\th)}{m_a m_b \sinh\th} \,\,,
\lab{ds}
\EN
where 
\EQ
F^\Psi_{\bar{c}\bar{d}ab}(\th_1-\th_2) \equiv
F^\Psi_{\bar{c}\bar{d}ab}(\th_1+i\pi,\th_2+i\pi,\th_1,\th_2) \,\,\,.
\label{special}
\EN
The right hand side of (\ref{ds}) employes the expression 
of the Form Factor at very special values of the rapidity variables. 
According to eq.\,(\ref{kinpole}), the Form Factors present pole 
singularities whenever the rapidities of a particle-antiparticle pair 
differ by $i \pi$ and, in fact, these kinematical poles are 
often explicitly inserted into the denominator of their parameterization. 
Apart from a term encoding the monodromy properties, this 
parameterization may be written as ${\cal Q}/{\cal D}$ where 
both ${\cal Q}$ and ${\cal D}$ are polynomials in the variables 
$\cosh\theta_{ij}$, the denominator being fixed by the pole structure whereas
the numerator determined by means of the residue equations, as 
for instance those of eq.\,(\ref{kinpole}). From the finiteness of the 
left hand side of eq.\,(\ref{ds}), we expect therefore that the 
``$i \pi$ singularities" of the denominator of the Form Factors 
$F^\Psi_{\bar{a}\bar{b}ab}(\th_1,\th_2,\th_3,\th_4)$ should be cancelled 
by the polynomial ${\cal Q}$, once evaluated at the specific rapidity 
configuration of eq.\,(\ref{special}). This prediction will be 
explicitly checked in the next section and should hold in general
whenever the perturbing operator is local with respect to the fields which
create the particles in the unperturbed theory\footnote{We defer to the
section devoted to the Ising model the discussion of the case in which this 
condition is not fulfilled.}. 

Eqs.\,(\ref{dm}), (\ref{de}) and (\ref{ds}) are the main results of this 
section. The best use of these formulas is to get rid of the explicit 
dependence on the normalisation of the perturbing operator by defining 
universal quantities, as for instance ratios of the mass shifts. 
Hence, under the validity of the linear approximation, all the universal 
quantities of non-integrable field theories can be entirely expressed in 
terms of the Form Factors of the integrable ones. Comparison of the theoretical 
predictions with their numerical determinations will be presented in the 
next sections of the paper.  

It is particularly instructive to specialise the above 
discussion to the ``trivial'' case in which the perturbing operator 
$\Psi(x)$ coincides with the operator $\Phi(x)$ that defines the initial 
integrable theory. In this case of course the physics should be invariant 
since the result of the additional perturbation simply corresponds to a shift 
of the coupling constant of the original integrable model by an amount 
$\delta g=\l$. The variations of the masses of the particles and of the vacuum 
energy density corresponding to such a shift can be directly computed from 
eqs.\,(\ref{ma}) and (\ref{evac}), respectively. On the other hand, we can 
also apply our general formulas (\ref{dm}) and (\ref{de}) to estimate 
the first order corrections. The two different routes coincide as long 
as the following identities are valid 
\EQ
\begin{array}{l}
F_{\bar{a}a}^{\Theta}(i\pi,0)=2\pi m_a^2\,\,, \\
{\cal E}_{\rm vac}=\frac{1}{4\pi}\langle0|\Theta|0\rangle\,\,,
\end{array}
\label{consi}
\EN
where $\Theta(x)=2\pi g(2-x_\Phi)\,\Phi(x)$ is the trace of the energy-momentum
tensor for the trajectory $\chi=0$. The above two relationships are indeed 
true and can be easily derived by other means. However, it is interesting 
to notice that their validity emerges in this context for the role of 
consistency equations which they play. Obviously, 
by the same token we can generate an infinite number of identities
involving Form Factors of the original integrable field theory by 
considering higher multi--particle scattering processes. For 
instance, next to (\ref{consi}), a new identity is obtained by comparing 
eq.\,(\ref{total}) with eq.\,(\ref{ds}): since $\chi$ is constant in the case 
we are considering, we have 
\EQ
\frac{\partial S_{ab}^{cd}(\th)}{\partial\th}=-\frac{1}{2\pi
i}\frac{F^\Theta_{\bar{c}\bar{d}ab}(\th)}{s_{ab}(\th)}\,\,\,.
\lab{derivative}
\EN
This identity provides a simple and unique way to normalise the 
four-particle Form Factors of the stress-energy tensor. It may be then 
particularly useful in the study of massless field theories where the 
first relationship in eq.\,(\ref{consi}) cannot be used for this 
purpose. For instance, it is easy to check that eq.\,({\ref{derivative}) 
applies to the Form Factor of the massless model considered in 
ref.\,\cite{DMSmass}.

It is also obvious that the first order inelastic amplitudes
computable by formula (\ref{firstorder}) must vanish identically when we 
choose $\Psi(x)=\Phi(x)$. This is ensured by the fact that the form factors
of the stress-energy tensor $F^\Theta_{a_1\ldots a_n}(\th_1,\ldots,\th_n)$ 
factorise the term $P_\mu P^\mu$, with $P^\mu=\sum_{i=1}^n p_i^\mu$ denoting
the total energy-momentum of the set of particles. Since $p_i^\mu\to
-p_i^\mu$ when the $i$--th particle is crossed from the initial to the
final state, $P^\mu$ is zero for a set of particles entering a physical
scattering process. Only in the case of elastic scattering, the zeros
coming from the factor $P_\mu P^\mu$ are cancelled by the kinematical
poles and therefore relations analogous to eq.\,(\ref{derivative}) are 
obtained.

\resection{Non-Integrable Deformations of the Minimal Model $M_{(2,7)}$}

As already indicated in the introduction, the main idea of using a 
perturbative expansion based on the Form Factors of integrable theories 
is because we expect that this kind of series should be capable of 
approximating the dynamics of the non-integrable field theories close to 
the integrable ones better than any other approach. One of the physical 
reasons for this expectation is that the integrable field theories should 
provide from the start the right multi-particle basis in the Hilbert space 
of the perturbed, non-integrable ones: hence, the differences between 
their physical properties are presumed to be small and calculable. 
Said in mathematical terms, such perturbative expansion should be 
particularly significant since the exact expression of the Form Factor 
already corresponds to a resummation of an infinite number of terms 
originating from the action (\ref{multiple}). However, the final conclusion 
on the efficiency of the Form Factor perturbative approach should come from 
some explicit and direct comparison with a set of data obtained from  
other sources. This is what we are going to accomplish in this section as 
well as in the next one. As our first example, we will discuss the scaling 
region around the fixed point described by the non-unitarity minimal model 
$M_{(2,7)}$. This model is particularly appealing for its simplified 
dynamics whereby the significant physical effects we are looking for 
will not be masked by other additional complications.  

The minimal conformal model $M_{(2,7)}$ has only two primary operators, 
$\Phi_{1,3}$ and $\Phi_{1,2}$, both of them relevant with scaling 
dimensions given by $-6/7$ and $-4/7$ respectively \cite{BPZ}. 
The perturbations of the conformal action either by the operator $\Phi_{1,3}$ 
or by the operator $\Phi_{1,2}$ are both known to be, separately, integrable
\cite{FKM,KMM}. In their massive phase, both perturbations have two stable 
massive particles denoted by $A_1(\theta)$ and $A_2(\theta)$, with a mass 
ratio and a scattering matrix which depend, however, on the integrable 
direction considered. The exact two-particle elastic $S$-matrix and other 
relevant information about the two different integrable deformations are 
summarized in the following table
\EQ
\begin{tabular}{|c|c|} \hline
$\Phi_{1,3}$ deformation & $\Phi_{1,2}$ deformation \\ \hline
$\begin{array}{ccl}
S_{11}(\th) & = & \st{\bf 2}{\left(\frac{2}{5}\right)} \\
S_{12}(\th) & = & \st{\bf 2}{\left(\frac{3}{5}\right)} \,
\st{\bf 1}{\left(\frac{4}{5}\right)}  \\
S_{22}(\th) & = & \st{\bf 1}{\left(\frac{4}{5}\right)}
\,\left(\frac{2}{5}\right)^2
\end{array}
$  
&
$ 
\begin{array}{ccl}
S_{11}(\th) & = & \st{\bf 1}{\left(\frac{2}{3}\right)} \,
\st{\bf 2}{\left(\frac{1}{9}\right)} \,\left(-\frac{2}{9}\right)  \\
S_{12}(\th) & = & \st{\bf 1}{\left(\frac{17}{18}\right)}\,
\left(\frac{11}{18}\right) \\
S_{22}(\th) & = & \st{\bf 2}{\left(\frac{2}{3}\right)}
\,\left(\frac{1}{9}\right) \, \left(\frac{5}{9}\right) 
\end{array}
$ 
\\ \hline
$
\frac{m_2}{m_1} = 2\,\cos\frac{\pi}{5} = 1.6180...
$
& 
$
\frac{m_2}{m_1} = 2\,\cos\frac{\pi}{18} = 1.9696... 
$ \\
\hline
$ {\cal E}_{vac} = - \frac{m_1^2}{8 \sin \frac{2\pi}{5}} = 
- 0.1314.. m^2_1 $ 
& 
$ {\cal E}_{vac} = - \frac{m_1^2}{8\left(\sin\frac{\pi}{3} 
+ \sin\frac{\pi}{9} - \sin\frac{2\pi}{9}\right)} = 
-0.2211.. m^2_1 $\\
\hline
\end{tabular}
\label{information}
\EN
where 
\[
(\al) \equiv  
\frac{\tanh\frac{1}{2}(\th + i \pi\al)}{\tanh\frac{1}{2}(\th - i \pi\al)}
\,\, ,
\]
and the bound state poles in the $S$-matrix amplitudes related to the 
particles $A_i$ are identified by the index $i$ placed above the functions 
$(\al)$. 

At the present, the $\Phi_{1,3}$ perturbation is the most studied of the 
two and in particular the Form Factors of the two primary fields are 
known \cite{koubek}. This suggests to adopt as our initial euclidean 
action ${\cal A}_0$ that relative to the field $\Phi_{1,3}$, i.e. 
\EQ
{\cal A}_0 = {\cal A}_{(2,7)} + g \int d^2x\, \Phi_{1,3}(x) \,\, ,
\EN
and then add to it the other deformation \footnote{Throughout this 
section we will refer to the region of the
coupling space of the theory (\ref{action27}) in which, within the
standard CFT normalisation of the fields, $g$ is real and
negative and $\lambda$ is purely imaginary with positive imaginary
part. The presence of the imaginary 
coupling is not surprising in view of the non--unitary nature of 
the original conformal model.}
\EQ
{\cal A} = {\cal A}_0 + \l \int d^2x\, \Phi_{1,2}(x)\,\,\,.
\lab{action27}
\EN
It is natural to assume that the spectrum associated to the action 
(\ref{action27}) consists for all positive values of the dimensionless 
parameter $\chi$ of two massive non-degenerate excitations (the validity of 
this assumption can be directly confirmed by means of a numerical technique 
that we will mention below). Hence, the QFT described by (\ref{action27}) is 
of the type of 
isospectral theory discussed in Section 2. This circumstance, together
with the relative simplicity of the scattering theory in both integrable 
directions, makes this theory the ideal playground for testing the 
Form Factor perturbative scheme. In order to make explicit predictions,  
we have listed in Appendix C the matrix elements of the operator 
$\Phi_{1,2}$ that we need in the sequel. 

Let us initially compute the first order correction to the elastic scattering 
amplitude $S_{11}(\th)$. According to the general formula (\ref{ds}), this is
given in terms of the four-particle Form Factor given by eq.\,(\ref{fourff}), 
computed at specific values of the rapidities. The corresponding 
particle configuration consists of neutral particles with rapidities 
differing by $i\pi$, i.e. the FF is on the resonant configuration ruled  
by the ``kinematical poles'' of the factors $\cosh(\th_{kl}/2)$ in the 
denominator of (\ref{fourff}). However, the factor $\Q(\th_{ij})$ present 
in the numerator of the four-particle Form Factor exactly cancels these 
divergences and produces as a finite result
\EQ
S_{11}(\th,\lambda) \simeq 
\frac{\tanh\frac{1}{2}\left(\th + i \frac{2\pi}{5}\right)}
{\tanh\frac{1}{2}\left(\th - i\frac{2\pi}{5}\right)}
- i \frac{\l c_0}{m_1(g,0)^2} 
\left(32 \sin^2\frac{\pi}{5} \right)
\frac{\cosh^2\frac{\th}{2}}{\sinh\th}\,\,
\frac{1+ 2\cos\frac{2\pi}{5} \cosh\th}
{(\sinh\th - i\sin\frac{2\pi}{5})^2}\,\,.
\lab{ds27}
\EN
The rapidities employed in the above formula are obviously defined on the 
unperturbed mass-shell condition. The finiteness of the above formula is an 
explicit demonstration of the cancellation of the kinematical singularities of 
the FF discussed in the previous section. 

The correction term in (\ref{ds27}) exhibits a second order pole at
$\th=2\pi/5$. The increased order of the pole signals that the position 
of the bound state pole in the $\th$-plane has become in the non-integrable 
theory a function of the coupling constant $\lambda$. This suggests the 
reabsorbing of the first order term in (\ref{ds27}) as a correction to 
the location of the original simple pole, given now by 
\EQ
\th= i u = i\left(\frac{2\pi}{5} + 
\frac{\l c_0}{(m_1^0)^2} 8 (\cos\frac{2\pi}{5}+2\cos\frac{\pi}{5})\right) 
+ O(\l^2)\,\,\,.
\EN
For the mass of the heaviest particle we then have 
\bea
m_2 &=& 2 m_1^0 \cos\frac{u}{2}\nonumber \simeq \nonumber\\
      &\simeq & 2m_1^0 \cos\left[\frac{\pi}{5} 
\left(1 - \frac{\l c_0}{(m_1^0)^2} 8 \sin\frac{\pi}{5} 
\left(1+2\cos^2\frac{2\pi}{5}\right)\right)\right] = 
\\
& = &
m_2^0 
\left[1 - \frac{\l c_0}{(m_1^0)^2} 8 \sin\frac{\pi}{5} 
\left(1+2\cos^2\frac{2\pi}{5}\right)\right]
\,\, , \nonumber
\eea
where $m_i^0$ denote the unperturbed value of the masses. By using eq.\,(C.3) 
it is easy to check that the above expression indeed coincides with the value 
directly obtained from eq.\,(\ref{dm}), i.e. 
\EQ
m_2^2 = (m_2^0)^2 + 2 \l \,F_{22}(i\pi)\,\,.
\EN

In addition to this consistency check, a direct test of the theoretical 
predictions for the variations in the spectrum of the theory under the 
perturbation can be obtained by the so-called ``truncation method'' 
\cite{YZtrunc}. The basic idea of this approach is to study the theory 
on an infinitely long strip of width $R$ (the linear spatial volume) with 
periodic boundary conditions. After choosing a Hilbert space basis of 
eigenvectors of the ultraviolet conformal Hamiltonian, all the matrix 
elements of the perturbed Hamiltonian on this basis can be exactly computed. 
The off-critical spectrum can then be found by numerical diagonalisation
on a truncated Hilbert space containing a suitable number of
states\footnote{All these steps can be performed by means of the 
algorithm developped in \cite{LM}.}. Since the method does not rely in any way 
on integrability, it can be applied to any perturbation of the conformal 
point. In particular, for the theory defined by the action (\ref{action}), the
energy levels must have the scaling form
\EQ
E_i(R,g,\lambda)=\frac{2\pi}{R} 
f_i(m_1R;\chi)\,\,,\hspace{.8cm}i=0,1,2\ldots\,\,\,.
\EN
At very short distance scales, the critical fluctuations are expected
to dominate so that the spectrum must coincide with that of the
conformal point given by \cite{cardy}
\EQ
E_i\simeq\frac{2\pi}{R}\left(x_i-\frac{c}{12}\right)\,\,,\hspace{.8cm}
m_1R<<1\,\, ,
\EN
where $c$ denotes the central charge and $x_i$ the scaling dimensions
of the scaling fields in the underlying conformal theory. In the
infrared limit, on the other hand, one should recover the spectrum of the 
massive theory on the plane and therefore the energy levels are given 
by 
\EQ
E_i\simeq{\cal E}_{vac}(g,\chi)R+m_i(g,\chi)\,\,,\hspace{.8cm}m_1 R >> 1
\lab{spectrum}
\EN
where the first term takes into account the vacuum bulk energy
contribution and $m_i$ denotes the mass-gap of the i-th level\footnote{
The above asymptotic expression only applies in the ideal situation 
where we discard the effects induced by the truncation of the Hilbert space. 
Although one can also take into account these extra numerical effects and 
refine consequently the above formulas, the long practice with integrable 
massive perturbations has shown that as far as 
the low lying energy levels are concerned, they are not much affected 
by truncation effects so that they can be obtained with a remarkable 
accuracy on a large range of $R$ by just including few conformal states in 
the calculation. This observation obviously helps to speed up the numerical 
work.}.
The first energy levels in the spectrum of the pure $\Phi_{1,3}$
perturbation of the conformal model $M_{(2,7)}$ obtained by including all 
conformal states up to level five in the Verma modules are shown in
Fig. 3. Starting from the bottom, the first four parallel lines are 
easily identified as the vacuum energy level, the one-particle energy level 
of the lowest particle $A_1$, the one-particle energy level of the heaviest 
particle $A_2$ and the two-particle threshold energy line, respectively. 
The remaining levels will be part of the continuum in the infinite 
volume limit. The measured values of the mass ratio $m_2^0/m_1^0\simeq 
1.61$ and the vacuum energy density ${\cal E}^0_{vac}\simeq -0.13 (m_1^0)^2$ 
are in good agreement with the theoretical expectations of the corresponding  
integrable theory. As for other spectra of integrable models 
\cite{YZtrunc,LMC,LMart,Marcio}, several crossings of the energy levels are 
expected and they are indeed observed for this integrable deformation of 
the $M_{(2,7)}$ model. When a small $\Phi_{1,2}$ perturbation is switched on 
(keeping $g$ fixed), the first qualitative effect consists to resolve the 
degeneracy present at the crossing points, i.e. all energy levels have now the 
tendency to repel each other. This feature can be interpreted 
as the breaking of the initial integrability of the model since it is 
well known that, in absence of any higher symmetry, two hamiltonian lines 
cannot generally cross. At a quantitative 
level, for small values of $\lambda$, both the separation and the slope of 
the lines vary linearly with it. The first order variations of these 
quantities are measured to be
\bea
&& \frac{\delta m_2}{\delta m_1}\simeq 3.74\,\,, \nonumber \\
&& \frac{\delta{\cal E}_{vac}}{\delta m_1}\simeq 0.67\,\,m_1^0\,\,\,. \nonumber 
\eea
They must be compared with the theoretical predictions for the same
quantities obtained by using eqs. (\ref{dm}) and (\ref{de})
\bea
&& \frac{\delta m_2}{\delta m_1}=\frac{m_1^0}{m_2^0}
\frac{F^{(1,2)}_{22}(i\pi)}{F^{(1,2)}_{11}(i\pi)}=
2\left(4\cos\frac{\pi}{5} \sin^2 \frac{2\pi}{5} -1\right)=3.8541..
\,\,, \nonumber \\
&& \frac{\delta{\cal E}_{vac}}{\delta m_1} \,=\, m_1^0 \,
\frac{\langle 0|\phi_{1,2}|0\rangle}{F^{(1,2)}_{11}(i\pi)}=
-\frac{\cos\frac{\pi}{5}}{2\sin\frac{\pi}{5}}=0.68819..\, m_1^0\,\,\,.
\nonumber 
\eea
The agreement between the theoretical estimates and the 
measured values is therefore quite satisfactory, the discrepancy being 
of the same order of the numerical error introduced by truncation effects. 

The truncation method obviously allows us to study the spectrum of the theory 
for any values of the coupling constants $g$ and $\lambda$ and to 
explicitly test several theoretical assumptions. We have for instance 
verified the validity of the scaling law (\ref{ma}) for the spectrum, which 
in this theory always consists of two distinct one-particle states whose mass 
ratio smoothly interpolates between the values $m_2/m_1 = 
2 \cos\frac{\pi}{5}$ and $m_2/m_1 = 2\cos\frac{\pi}{18}$ (relative to the 
limits $\chi \to 0$ and $\chi \to \infty$, respectively), as shown in 
Figure 4. Obviously for large values of $\chi$ the physical 
properties of the model can be no longer theoretically predicted  
by the first terms of perturbative series based on the $\Phi_{1,3}$ 
deformation. In particular, to study the model in the limit $\chi\to\infty$,
it is evident that 
instead of including higher order terms of the perturbative series based 
on the $\Phi_{1,3}$ deformation, it would be more convenient to make 
use of the other perturbation theory based on the second integrable 
operator $\Phi_{1,2}$ deformation. 

\resection{The Scaling Region of the Two--Dimensional Ising Model}

Aim of this section is to extract through the Form Factor perturbative 
techniques some information about the scaling region nearby the critical 
point of the two-dimensional Ising model described by 
\EQ
{\cal A} = {\cal A}_{(3,4)} + \tau \int d^2x\,
\varepsilon(x) + h \int d^2x\,\sigma(x)\,\, ,
\lab{ising}
\EN
and also to study the model by means of the truncation method. 
As expressed by (\ref{ising}), the most general off--critical 
realization of the Ising model is given by a perturbation of the 
simplest unitary minimal model $M_{(3,4)}$ with a linear combination of 
its two only relevant primary fields, i.e. the energy density 
$\Phi_{1,3}(x) \equiv \varepsilon(x)$ of scaling dimension
$x_\varepsilon=1$ and the magnetization $\Phi_{1,2}(x) \equiv \sigma(x)$ of
scaling dimension $x_\sigma=1/8$. The conjugated couplings $\tau$ and
$h$ are respectively interpreted as the deviation from the critical
temperature $T_c$ and a constant magnetic field, with physical dimensions 
given by $\tau\simeq M$ and $h\simeq M^{15/8}$, where $M$ is a mass scale. 
The Zamolodchikov's counting argument \cite{Zam} ensures that the theory 
(\ref{ising}) becomes integrable as far as one of the two coupling constants 
$\tau$ or $h$ is set equal to zero. The integrability of the purely 
thermal perturbation has been known for long time and can be 
neatly reformulated in terms of field theory of Majorana fermions 
\cite{KW,YZ,Ising1,McBook,Ising3}. The integrability of the magnetic 
deformation has been established by Zamolodchikov \cite{Zam} in the recent 
past and, for the original features of his findings this result may 
be regarded as a direct success of the application of field theoretical 
methods to statistical mechanics.

The action (\ref{ising}) defines a one parameter family of field theories 
which can be labelled by the dimensionless combination
$\chi\equiv \tau|h|^{-8/15}\in(-\infty,+\infty)$. Previous investigations 
have shown that the particle content of the model 
must undergo drastic changes as a function of $\chi$ \cite{mccoy-wu,canberra}.
As we will see in the following, this prediction can be directly confirmed by 
the numerical determination of the spectrum for different values of $\chi$
(see Figures 7). The reason for such behaviour of the spectrum 
will be easily understood after reviewing some exact results obtained 
for the two integrable directions.

\subsection{Thermal Deformation}

Consider first the case $h=0$. In the high temperature phase, i.e. 
$\chi=+\infty$, the theory has an unique vacuum state and the spectrum 
consists of a single particle $A$ of mass $m = 2\pi \tau$ and multi-particle 
states theoreof. This is clearly confirmed by the truncation method 
calculation shown in Figure (7.a): starting from the bottom, the sequence of 
energy levels is given by the ground state, one-particle level and then all 
the multi--particle lines of the continuum, the lowest of which corresponding 
to the two-particle threshold. 
The massive excitation present in this phase of the model can be 
equivalently regarded as a free neutral fermion or as an interacting boson 
(with scattering amplitude $S=-1$) created by the magnetization operator. 
Invariance under the spin reversal implies that the ``order'' (``disorder'') 
field $\sigma(x)$ ($\mu(x)$) couples to the states with an odd (even) number 
of particles only. The Form Factors for the two operators are given by the 
unique expression \cite{KW,YZ}
\EQ
F_{(n)}(\th_1,\ldots,\th_n)\equiv\langle
0|{\cal O}(0)|a(\th_1)\ldots
a(\th_n)\rangle \,=\, (i)^{\left[\frac{n}{2} -1\right]} 
\prod_{i<j}^n \tanh\frac{\th_i-\th_j}{2}\,\,,
\lab{sigmaff}
\EN
with $n$ odd for ${\cal O}(x) = \sigma(x)$ and even for ${\cal O}(x) = 
\mu(x)$ and $[\alpha]$ denoting the integer part of $\alpha$. 

The situation in the low temperature phase, i.e. $\chi=-\infty$, is easily
deduced from the previous one by duality. In fact, the excitations are now 
created by the disorder operator $\mu(x)$ and must be interpreted as kinks 
and anti--kinks of mass $m$ interpolating between the two degenerate vacua 
coming from the spontaneous symmetry breaking of the $Z_2$ invariance 
of the model. The Form Factors of $\sigma(x)$ coincide with those of the 
operator $\mu(x)$ computed in the high-temperature phase $T>T_c$. Hence, 
the magnetization operator $\sigma(x)$ now couples only to states with even 
number of excitations. The numerical determination of the spectrum is shown 
in Figure (7.g). The lowest two lines are clearly the degenerate ground 
state levels. They approach each other exponentially in the crossover 
region, $\Delta E \sim \exp(-m R)$, where $m$ is the mass of the kink 
responsible of the finite-size tunnelling effect between the two vacua. 
Since periodic boundary conditions were adopted on 
the strip of width $R$ in the numerical determination of the spectrum, 
in Figure (7.g) there are no energy levels corresponding 
to odd number of kink states. In particular, the energy line associated 
to the single kink state is absent and therefore the lowest energy level  
above the ground state energies is given in this case by the  
kink-antikink threshold line. 

\subsection{Magnetic Deformation} 

Far richer is the situation for the other integrable direction 
obtained when $\tau=0$ and $h\neq 0$, i.e. $\chi=0$. In fact, Zamolodchikov 
has shown that in this case the spectrum consists of eight different species 
of self-conjugated particles $A_{i}$, $i=1,\ldots,8$ with masses\footnote{
Within the standard CFT normalization of the magnetization operator,
obtained  by the equation
$
<\sigma(x)\sigma(0)>=\frac{1}{\,\,|x|^{\frac{1}{4}}}
$
($ |x|\goto 0$), the overall mass scale $M(h)$ has been exactly determined 
in \cite{TBA},
$
M(h) \,=\, {\cal C} \, h^{\frac{8}{15}}
$ 
where
\[
{\cal C} \,=\,
 \frac{4 \sin\frac{\pi}{5} \Gamma\left(\frac{1}{5}\right)}
{\Gamma\left(\frac{2}{3}\right) \Gamma\left(\frac{8}{15}\right)}
\left(\frac{4 \pi^2 \Gamma\left(\frac{3}{4}\right)
\Gamma^2\left(\frac{13}{16}\right)}{\Gamma\left(\frac{1}{4}\right)
\Gamma^2\left(\frac{3}{16}\right)}\right)^{\frac{4}{5}} \,
 \,=\, 4.40490858...
\]
} \cite{Zam}
\bea
m_1 &=& M(h)\,, \nonumber \\
m_2 &=& 2 m_1 \cos\frac{\pi}{5} = (1.6180..) \,m_1\,,\nonumber\\
m_3 &=& 2 m_1 \cos\frac{\pi}{30} = (1.9890..) \,m_1\,,\nonumber\\
m_4 &=& 2 m_2 \cos\frac{7\pi}{30} = (2.4048..) \,m_1\,,\nonumber \\
m_5 &=& 2 m_2 \cos\frac{2\pi}{15} = (2.9562..) \,m_1\,,\\
m_6 &=& 2 m_2 \cos\frac{\pi}{30} = (3.2183..) \,m_1\,,\nonumber\\
m_7 &=& 4 m_2 \cos\frac{\pi}{5}\cos\frac{7\pi}{30} = (3.8911..) \,m_1\,,
\nonumber\\
m_8 &=& 4 m_2 \cos\frac{\pi}{5}\cos\frac{2\pi}{15} = (4.7833..) \,m_1\,,
\nonumber
\eea
Notice that only the first three particles of the above spectrum lie below 
the lowest threshold at $2m_1$. The remaining particles are however stable 
since integrability prevents the possibility of inelastic processes, in 
particular decay processes. The $S$-matrix describing the interactions 
between the eight particles was exactly determined in \cite{Zam}. These 
results were used in refs.\,\cite{DM} to implement the bootstrap 
equations for the Form Factors of the model and to compute in particular  
the one and two-particle matrix elements of the operator $\sigma(x)$. As 
expected from the fact that neither the action nor the $S$-matrix of the model 
exhibit any internal symmetry, the magnetization operator was found to couple 
to all the eight particles in the spectrum, namely 
$F_i^\sigma=\langle 0|\sigma|A_i\rangle\neq 0$ for $i=1,\ldots,8$. 

The finite size spectrum of the Ising model in a magnetic field at
$T=T_c$ as obtained with the truncation method is shown in
Figure (7.d). Starting from the bottom, five parallel lines are clearly
identified in the large volume limit. They correspond to the ground
state, the three lightest particles and the first two-particle
threshold, respectively. The remaining single particle states are of more 
difficult identification since they are placed among the higher threshold 
lines. Notice that they cross in several points the
``momentum lines'' (i.e. lines corresponding to states containing two
or more particles not at rest with respect to each other) converging
toward the threshold in the infinite volume limit. Level crossing in
the energy spectrum is always related to the presence of symmetries
and in the present case strongly supports the integrability of the theory.

\subsection{McCoy--Wu Scenario}

After having discussed individually each deformation, let us consider the more 
general case described by the action (\ref{ising}). Following McCoy and Wu 
\cite{mccoy-wu}, let us introduce the Fourier transform of the spin-spin 
correlation function
\EQ
G(p^2,\chi)=\int d^2x\,e^{ipx}\langle\sigma(x)\sigma(0)\rangle\,\,\,.
\nonumber
\EN
According to the discussion of the previous subsections, the leading infrared 
singularities of this function in the complex $p$-plane for $\chi=+\infty$, 
$0$, $-\infty$ are those shown in Figs.\,(6.a), (6.d) and (6.g), respectively. 
Since these three situations correspond to different values of the
couplings in the same QFT defined by the action (\ref{ising}),
it must be possible to interpolate continously between 
them by moving along the semicircular path 
drawn in Figure 5. To describe the change of physical properties of the theory 
along the path, the following scenario has been proposed
\cite{mccoy-wu} : starting from the point $a$ in Figure 5, the spin 
reversal symmetry will be broken as far as a magnetic field $h$ is turned on 
and consequently, even particle thresholds immediately appear, as shown in
Fig. 6.b. Moreover moving toward the point $d$, additional poles
emerge from unphysical sheets of the Riemann surface through the branch 
cuts and become physical bound state poles. From Zamolodchikov's solution, 
we know that at the point $d$ there are precisely three such poles below the 
first two-particle threshold. Their number however continues to increase when 
moving toward the negative $\tau$ axis until altogether they coalesce at 
point $g$ and give rise to the branch cut starting at $2im$ in Fig.\,(6.g). 
The lowest lines of the numerical spectra corresponding to the values
\footnote{The quoted values of $\chi$ refer to the standard CFT
normalisation of the fields $\varepsilon(x)$ and $\sigma(x)$. For a
generic field $\varphi$ of scaling dimension $x$, this is defined
by the condition $\langle\varphi(r)\varphi(0)\rangle\goto r^{-2x}$,
$r\goto 0$.} $\chi=+\infty$, $1.32$, $0.16$, $0$, $-0.16$, $-1.32$, 
$-\infty$ are shown in the sequence of Figures 7 and we will comment 
on them below. 

The coalescence of the poles is quite a striking phenomenon. It was 
quantitatively discussed in ref.\,\cite{mccoy-wu} where the knowledge of the 
$n$-point correlators for the purely thermal Ising model was
exploited to study the behaviour at small $h$ of the spin-spin
correlation function expressed in the form
\EQ
\langle\sigma(x)\sigma(0)\rangle_{conn,\,\,h}=\sum_{n=0}^\infty\frac{1}{n!}h^n
\int d^2x_1\ldots d^2x_n\langle\sigma(x)\sigma(0)\sigma(x_1)\ldots 
\sigma(x_n)\rangle_{conn,\,\,h=0}\,\,\,.
\nonumber
\EN
It was found that when the magnetic field is switched on, the
two-particle branch cut of Fig.\,(6.g) breaks up in a sequence of poles
located at
\EQ
(2 + h^{2/3} \gamma_k^{2/3})im \,\, ,
\label{chain}
\EN
where $\gamma_k$ are the positive solutions of 
\[
J_{\frac{1}{3}} \left(\frac{1}{3} \gamma_k\right) + 
J_{-\frac{1}{3}} \left(\frac{1}{3} \gamma_k\right) = 0 \,\, ,
\]
and $J_{\nu}(x)$ is the Bessel function of order $\nu$. The above values 
coincide with the eigenvalues of the Schroedinger equation for a particle 
in a central linear potential. The physical origin of this result is that 
the presence of a small magnetic field breaks the degeneracy 
of the two vacua in the low-temperature phase inducing a linear confining  
potential between the kinks\footnote{One can easily convince himself of this 
feature by using a semi-classical analysis of the kink states.}. It is clearly 
evident then that the magnetic field induces a drastic change in 
the structure of 
the low-temperature phase of the model: in fact, as soon as the magnetic 
field is switched on, the kink configurations are no longer asymptotic 
states of the field theory and the original kink--antikink pairs collapse 
into a sequence of bound states. This happens for {\em any} value of the 
magnetic field $h$, however small. The lightest of these bound states has  
a mass equal to $2m + O(h^{2/3})$, so that the first branch point 
is now located at $4im + O(h^{2/3})$.

\subsection{Form Factor Perturbation Theory}

Let us now turn to the discussion of the Ising model within the
perturbative framework considered in this paper
\footnote{For other perturbative studies of the Ising model field
theory see refs.\,\cite{McCoy,Dotsenko}.}. 
Our main interest will 
be to develop the perturbative picture around the integrable magnetic 
direction ($\chi=0$), for which no other approach is presently avalaible.
Before doing that, we will however briefly comment on the case $h\sim 0$, 
distinguishing the two situations $T > T_c$ and $T < T_c$. 

At $T>T_c$, the magnetization operator $\sigma(x)$ couples to the 
odd-particle states only. Hence, the opening of inelastic channels 
(e.g. the production process $A A \goto A A A$) with the consequent 
breakdown of integrability can be easily checked at first order in $h$ but 
the correction to the mass or the variation of the elastic scattering 
amplitudes occurs instead only at the second order in $h$. The breaking of 
the integrability implies that the numerical spectrum no longer presents in 
this case crossing of the energy levels, a feature which, although not
always clearly visible in Figs.\,7, is however 
confirmed by the numerical data. 

In the low-temperature phase $T < T_c$ 
the spectrum of the theory is known to undergo a qualitative and 
non-analytic change when the magnetic field is switched on. Although 
these considerations naturally suggest the failure of any finite order 
of a perturbative approach to study this phenomenon,  it is nevertheless 
instructive to see how such failure manifests itself in our method. 
To this aim, it is sufficient to attempt the computation of the correction to 
the mass of the kink, an excitation which we know no longer exists 
as an asymptotic state in the perturbed theory. According to the general
formula (\ref{dm}), the first order correction is proportional to the
two-particle form factor of the magnetization operator computed for a
rapidity difference equal to $i\pi$. Since this form factor is equal 
to $\tanh\frac{\th_1-\th_2}{2}$ (see eq.\,(\ref{sigmaff})), it has a
pole at $\th_1-\th_2=i\pi$ and therefore an unbounded correction is 
obtained for the mass of the kink. 
The general theory of Form Factor predicts that such ``kinematical" pole 
is present in the two-particle Form Factor $\langle 0|{\cal
O}(0)|A(\th_1)\bar{A}(\th_2)\rangle$ only if the operator 
${\cal O}(x)$ is non-local with respect to the field which creates the
particle $A$ (as it is the case for the fields $\sigma(x)$ and $\mu(x)$
in the Ising model). This circumstance suggests therefore that a confinement
phenomenon of the type described above for the Ising model has to be
expected each time that the perturbing operator and the fields which create 
the particle excitations in the unperturbed theory are not mutually
local. It would be interesting to check this prediction in other statistical 
models where the integrable spectrum is given by kink excitations. 

In the case of the Ising model, the numerical determination of the spectrum 
confirms the above scenario. From Figure (7.f), we see in fact that the first 
effect of 
$h$ consists in the removal of the degeneracy of the lowest as well as of 
the higher eigenvalues. The two originally degenerate ground state levels 
have now been splitted into: (i) a unique ground state energy line and 
(ii) an excited state whose mass gap diverges in the large volume limit 
$R \goto \infty$. The degeneracy of the lowest threshold line has also 
been lifted giving rise to a sequence of one-particle energy levels, the 
lowest values of which can be checked to be in reasonable agreement with 
eq.\,(\ref{chain}). By increasing the magnetic field $h$, the above 
characteristic feature of the spectrum are futher enhanced, as seen by 
following Figs.\,(7.e) and (7.f) in reverse order: the 
divergent energy line of the initial degenerate ground states meets all other 
lines at smaller values of $R$ (and therefore it decouples faster from 
the remaining spectrum) whereas the other eigenvalues start to assume 
the structure and the values predicted by the Zamolodchikov solution for 
$T = T_c$, reproduced in Figure (7.d). Observe that, while the
eigenvalues have 
the typical repulsive behaviour of a non-integrable situation all along the 
path from the point $f$ to $e$, on the contrary they cross each other once we 
have reached the integrable situation of the pure magnetic axis. Moreover, 
the fact that in the pure magnetic case the approach to the asymptotic value 
of the masses is reached from below can be simply interpreted as 
the reminiscence of the divergent line coming from the degenerate ground 
state energies of the low-temperature phase. 
 
Let us now analyse more closely and in quantitative terms the field theory 
defined in the vicinity of the magnetic axis.  
Perturbation theory around the magnetic axis ($T=T_c$ with $h\neq 0$) 
obviously requires the knowledge of the Form Factors of the energy 
operator $\varepsilon(x)$ in the Zamolodchikov field theory involving 
the eight massive states $A_i$ ($i=1,\cdots,8$) associated to the pure 
magnetic model. These Form Factors can be exactly computed by using the 
same bootstrap approach used in ref.\,\cite{DM} for the magnetization 
operator. The detailed discussion on the computations of these quantities 
will be presented elsewhere \cite{preparation2} and here we only quote the  
results needed for our present purposes
\bea
&& \langle 0|\varepsilon(0)|0\rangle=m_1\,\,,\nonumber\\
&& F_{11}^\varepsilon(i\pi)=\langle
0|\varepsilon(0)|A_1(\th+i\pi)A_1(\th)\rangle=-17.8933..m_1\,\,,\nonumber\\
&& F_{22}^\varepsilon(i\pi)=\langle
0|\varepsilon(0)|A_2(\th+i\pi)A_2(\th)\rangle=-24.9467..m_1\,\,,\nonumber\\
&& F_{33}^\varepsilon(i\pi)=\langle
0|\varepsilon(0)|A_3(\th+i\pi)A_3(\th)\rangle=-53.6799..m_1\,\,,\nonumber\\
&& F_{44}^\varepsilon(i\pi)=\langle
0|\varepsilon(0)|A_4(\th+i\pi)A_4(\th)\rangle=-49.3169..m_1\,\,\,.
\lab{energyff}
\eea
The first line of the above relationships should be meant as expressing 
the normalisation of the energy operator $\varepsilon(x)$.  
When the energy perturbation is switched on, the spectrum in Fig.\,(7.d)
undergoes continous deformations which can be followed by the
truncation method. In particular, for $|\tau|\sim|T-T_c|$ sufficiently
small and $h$ fixed, the variations in the energy levels are linear in
$\tau$ and their direct measurement with the truncation method 
is given by 
\bea
&& \frac{\delta{\cal E}_{vac}}{\delta m_1}\simeq -0.05\,m_1^0\,\,,\nonumber\\
&& \frac{\delta m_2}{\delta m_1}\simeq 0.87\,\,, \label{measure}\\
&& \frac{\delta m_3}{\delta m_1}\simeq 1.50\,\,\,.\nonumber
\eea
The theoretical predictions for the same quantities are obtained
plugging the values (\ref{energyff}) into the first order formulae
(\ref{de}) and (\ref{dm})
\bea
&& \frac{\delta{\cal E}_{vac}}{\delta m_1}=
\frac{\langle 0|\varepsilon|0\rangle}
{F_{11}^\varepsilon(i\pi)}\,m_1^0=-0.0558..m_1^0\,\,,\nonumber\\
&& \frac{\delta m_2}{\delta m_1} = 
\frac{F_{22}^\varepsilon(i\pi)}{F_{11}^\varepsilon(i\pi)}
\,\frac{m_1^0}{m_2^0}=0.8616..\,\,,\label{th}\\
&& \frac{\delta m_3}{\delta m_1} = \frac{F_{33}^\varepsilon(i\pi)}
{F_{11}^\varepsilon(i\pi)}\,\frac{m_1^0}{m_3^0} = 1.5082..\,\,\,.\nonumber
\eea
The agreement between the theoretical and numerical estimates should be 
regarded as quite satisfactory. 

More dramatic is the effect of the perturbation on the five particles above 
the threshold. We have already mentioned that their stability at
$\tau=0$ must be considered a consequence of the integrability of the
theory which prevents any kind of inelastic process. They are then
expected to decay in the perturbed, non-integrable theory. This is
easily seen to be the case by analysing the effect of the perturbation on
the finite size spectrum of the model. Indeed, since integrability is lost
under the perturbation, level crossing in the spectrum is no longer
allowed at $\tau\neq 0$. This means that the energy levels are now
forced to repel each other at the former crossing points. Since each
line associated to a particle above threshold crossed an infinite number of 
momentum lines at $\tau=0$ (Fig.\,(8.a)), it immediately becomes a ``broken
line'' (almost) parallel to the threshold when the perturbation is
switched on (Fig.\,(8.b)). This is exactly the signature of unstable
particles in the finite volume; more precisely, the difference in the slopes 
of the broken line and the threshold is proportional to the 
width of the resonance \cite{Luscher}.

In perturbation theory, the decay of the particles above threshold
manifests itself through the appearence of a negative imaginary part
in the mass. This is an effect which occurs though at the second order 
in $\tau$. To be specific, consider the lightest unstable particle $A_4$. 
It follows from (\ref{dm}) and (\ref{energyff}) that the first order 
correction is real and is given by
\EQ
\frac{\delta m_4}{\delta m_1} = 
\frac{F_{44}^\varepsilon(i\pi)}{F_{11}^\varepsilon(i\pi)}
\,\frac{m_1^0}{m_4^0} = 1.1460..\,\,\,.\nonumber
\EN
The computation of the second order correction requires the sum over a
complete set of intermediate $n$--particle states and involves an
energy denominator. Expressing the latter in the usual form
\EQ
\frac{1}{E-E_n+i\varepsilon}={\mbox
P}\,\left(\frac{1}{E-E_n}\right) - i\pi\delta(E-E_n)\,\,,
\nonumber
\EN
one immediately concludes that the principal part gives rise to the
second order correction to the real part of the mass while the delta
function originates an imaginary part receiving a contribution only
from the intermediate states whose total energy-momentum equals that
of the external particle. For the particle $A_4$ we have 
\EQ
{\mbox Im}\,m_4^2\simeq -\frac{\tau^2}{2m_1^0m_4^0\sinh\th^*}|{}_0\langle 
A_4(0)|\varepsilon(0)|A_1(\th^*)A_1(-\th^*)\rangle_0|^2\,\,,
\EN
where $\th^*=|{\mbox arccosh}\frac{m_4^0}{2m_1^0}|$.

\resection{Conclusions}

In this paper, we have shown that once the complete dynamics 
of two-dimensional exactly solvable models is known, this can be also 
extremely useful for 
investigating the structure of the quantum field theories close to 
the integrable
directions. In particular, we have derived a perturbation theory for the 
non-integrable QFT based on the Form Factors of the exactly solvable 
relativistic models and we have considered the lowest term 
of the series in order to study the effects of a small perturbation 
which breaks the integrability of the unperturbed theory. We have 
discussed 
the variation of the mass spectrum, the shift in the vacuum energy density 
and the correction to the elastic part of the $S$-matrix, and we have 
successfully checked them against their numerical estimates obtained by the 
truncation method in two models with an underlying CFT, namely 
the minimal model $M_{(2,7)}$ and the Ising model. 

It is worth pointing out that, specialising the perturbative formulas 
to the trivial case of a parallel perturbation along the original 
integrable model, one may obtain as a by-product an infinite set of 
useful identities for the Form Factors of the stress-energy tensor of the 
original integrable field theory. While this appears to be as a convenient 
method of deriving them, at the same time it explains the reason of their 
validity because they can be regarded as consistency equations for the 
perturbative scheme built up on the Form Factors. Known features 
of the Form Factors of the integrable relativistic models are then 
deeply inter-related with the general structure of quantum field theories.   
 
The methods illustrated in this paper may apply of course to other interesting 
physical situations in addition to those considered here. 
We would like to mention, for instance, two non-integrable lagrangian 
field theories which would be interesting to analyse in terms of their 
Form Factor perturbative series. The first one is given by 
\EQ
L = \frac{1}{2} (\partial_{\mu} \varphi)^2 - \frac{m^2}{6 g^2} 
\left[\left(1-\frac{\lambda}{3}\right) (2 e^{g \varphi} + e^{-2 g \varphi}) + 
\lambda (e^{g \varphi} + e^{-g \varphi})\right] \,\,\,.
\label{BDSH}
\EN
By varying the parameter $\lambda$ in the interval $[0,3]$, we can interpolate 
between the Bullogh-Dodd and the Sinh-Gordon models. Both these field theories 
are separately integrable and the matrix elements of their local operators have 
been computed in \cite{Sh,BD}. Notice that the $Z_2$ symmetry of the 
Sinh-Gordon model is always broken along the entire interpolating 
trajectory and it is only recovered at the end point $\lambda = 3$. 

The second non-integrable lagrangian model is obtained by adding higher 
``harmonics" to the original Sinh-Gordon interaction, the simplest 
example of this class of non-integrable models being the so-called 
double Sinh-Gordon model
\EQ
L = \frac{1}{2} (\partial_{\mu} \varphi)^2 - \frac{m^2}{2 g^2} 
\left[(1 - 4\lambda) \cosh g\varphi + \lambda \cosh 2 g \varphi\right] \,\,\,.
\label{double}
\EN
By varying $\lambda$ in the interval $[0,\frac{1}{4}]$, we can 
interpolate in this case between the integrable field theories given by 
the Sinh-Gordon theory with coupling constant $g$ and $2 g$. 

Notice that for both examples (\ref{BDSH}) and (\ref{double}), the 
integrability of the two theories at the two extremes of the interpolation 
interval implies strong constraints for the ordinary perturbation theory 
based on the coupling constant $g$. It would be then interesting to learn 
more about those models by comparing the results obtained in ordinary 
Feynman diagram perturbation theory with those derived by the perturbative 
series based on the Form Factors.

\vspace{10mm} 
{\em Acknowledgements}. We would like to thank J.L. Cardy and
A. Schwimmer for useful discussions and suggestions. G.D. and 
P.S. were supported by EPSRC GR/J78044 and HEFCW grants, respectively. 

\newpage
\appendix

\appsection

A simple example of quantum mechanics will help to clarify the 
simultaneous relevance of all the inelastic thresholds in any energy 
interval.  
Consider a one-dimensional inelastic scattering process of a particle of 
mass $m$ which hits a target with $n+1$ internal states of increasing 
energies $E_i$ ($i=0,1,\ldots,n$), where the excited states 
$\mid i>$ ($i=1,\ldots,n$) are given by $\mid i> = a^{\dagger}_i \mid 0>$ 
($a_i^{\dagger}$ and $a_i$ denote the creation and annihilation 
operators, with standard commutation relations). The states of this system 
(target plus particle) are described by the wave function 
$\mid \Psi> = \sum_{i=0}^n a_i^{\dagger} \mid 0> \psi_i(x)$ and as 
Hamiltonian of the system we choose 
\EQ
H \,=\, \frac{p^2}{2m} + \sum_{i=0} ^n E_i a^\dagger_i a_i - 
\frac{h^2}{2m} \delta(x) \,\left[ U_0 \sum_{i=0}^n a_i^{\dagger} a_i + 
\sum_{i=1}^n U_i (a_0^{\dagger} a_i + a^{\dagger}_i a_0)\right]
\EN
$U_0$ rules the elastic transition amplitudes $\mid i> \rightarrow \mid i>$ 
whereas $U_i$ are related to the inelastic reactions 
$\mid 0> \leftrightarrow \mid i>$. For this toy-model, it is quite simple 
to determine the phase-shift $\delta_0$ and therefore the $S$-matrix 
element $S= e^{2 i \delta_0}$ relative to the elastic channel 
$\mid 0> \rightarrow \mid 0>$. This is given by the formula 
\EQ
\tan \delta_0 = \frac{U_0}{2k} - \sum_{l=1}^n 
\frac{U_l^2}{2k (U_0 + 2 i k_l)} \,\, ,
\label{explicit}
\EN
where 
\begin{eqnarray*}
k^2 & =& \frac{2m}{h} (E-E_0) \,\, ,\\
k^2_j & = & \frac{2m}{h} (E - E_j) \,\,\,.
\end{eqnarray*}
The phase shift $\delta_0$ is real when the energy $E$ is below the first 
threshold $E_1$ and complex above but it is worth to note that also in 
the elastic region $0 < E < E_1$, its value is determined by {\em all}
the inelastic parameters of the problems, i.e. $U_l$ and $k_l$.

\appsection

As it is well known, a very effective algebraic description of integrable
theories can be obtained in terms of the Faddeev-Zamolodchikov (FZ)
creation and annihilation operators, $Z_a^+(\th)$ and $Z_a(\th)$. They
satisfy the algebra
\bea
&& Z_a^+(\th_1)Z_b^+(\th_2)=S_{ab}^{cd}(\th_1-\th_2)\,Z_d^+(\th_2)Z_c^+(\th_1)
\nonumber \\
&& Z_a(\th_1)Z_b(\th_2)=S_{cd}^{ab}(\th_1-\th_2)\,Z_d(\th_2)Z_c(\th_1) 
\lab{FZ} \\
&& Z_a(\th_1)Z_b^+(\th_2)=S_{bc}^{da}(\th_2-\th_1)\,Z_d^+(\th_2)Z_c(\th_1)
+2\pi\,\delta_{ab}\,\delta(\th_1-\th_2) \nonumber 
\eea
which can be regarded as a generalisation of the canonical commutation
relations. The operator $Z_a(\th)$ annihilates the vacuum while
$Z_a^+(\th)$ creates a particle of tipe $a$ with rapidity $\th$. The
space of states is generated by
\bea
&& |a_1(\th_1)\ldots a_n(\th_n)\rangle\,=\,Z_{a_1}^+(\th_1)\ldots
Z_{a_n}^+(\th_n)|0\rangle\,\,, \nonumber \\
&& \langle a_n(\th_n)\ldots a_1(\th_1)|\,=\,\langle 0|Z_{a_n}(\th_n)\ldots
Z_{a_1}(\th_1)\,\,\,.
\lab{states} 
\eea
The {\em physical} asymptotic states can be selected through the
following ordering prescription over rapidities: the states
(\ref{states}) are ``In'' states if $\th_1>\th_2>\ldots>\th_n$, and
``Out'' states if $\th_1<\th_2<\ldots<\th_n$.

It is easily checked that the unitarity and Yang-Baxter equations for
the S-matrix can be obtained requiring respectively the consistency
under double application and the associativity of the FZ algebra
(\ref{FZ}). Concerning the remaining fundamental property of the
S-matrix, namely the crossing relation
\EQ
S_{ab}^{cd}(\th)=S_{\bar{d}a}^{\bar{b}c}(i\pi-\th)\,\,,
\lab{crossing}
\EN
let's formally perform in the last equation in (\ref{FZ}) the analytic
continuation $\th_2\goto\th_2+i\pi$. Then, substituting particles $b$ and 
$d$ with their antiparticles $\bar{b}$ and $\bar{d}$ and using 
eq.\, (\ref{crossing}), we obtain
\EQ
Z_a(\th_1)Z_{\bar{b}}^+(\th_2+i\pi)=S^{ab}_{cd}(\th)\,Z_{\bar{d}}^+(\th_2+i\pi)
Z_c(\th_1)\,\,,\hspace{.6cm}(\th_1\neq\th_2)\,\,\,.
\EN
Comparison with the second equation in (\ref{FZ}) suggests the
identification
\EQ
Z_{\bar{b}}^+(\th+i\pi)=-Z_b(\th)\,\,\,.
\lab{cros}
\EN 
The proportionality constant $-1$ has been chosen in order to fit the
canonical cases corresponding to $S_{ab}^{cd}(\th)=\pm 1$. Indeed we
recall that the Fourier decomposition of a free (e.g. bosonic) field
\EQ
\varphi(x)=\frac{1}{2\pi}\int\frac{dk^1}{k^0}\left[a(k)e^{-ikx}+a^+(k)
e^{ikx}\right] \,\,,
\EN
immediately leads to the representation
\bea
&& a(k)=-\frac{i}{2}\,\int
dx^1\,e^{ikx}
\buildrel \leftrightarrow \over \partial_0
\varphi(x)=f(k)\,\,,\nonumber \\
&& a^+(k)=\frac{i}{2}\,\int
dx^1\,e^{-ikx}
\buildrel \leftrightarrow \over \partial_0
\varphi(x)=-f(-k)\,\,\,.\nonumber 
\eea
Consider the two matrix elements 
\EQ
{}_{c_0} F^{\cal O}_{a_1\ldots a_n}(\th|\th_1,\ldots,\th_n)=
\langle 0|[Z_{c_0}(\th),{\cal
O}(0)]|a_1(\th_1),\ldots,a_n(\th_n)\rangle+
\langle 0|{\cal
O}(0)Z_{c_0}(\th)|a_1(\th_1),\ldots,a_n(\th_n)\rangle\,\,,
\EN
\EQ
F^{\cal O}_{\bar{c}_0a_1\ldots a_n}(\th,\th_1,\ldots,\th_n)=
\langle 0|[{\cal
O}(0),Z_{\bar{c}_0}^+]|a_1(\th_1),\ldots,a_n(\th_n)\rangle\,\,\,.
\EN
We can use the identification (\ref{cros}) and the FZ algebra to
recognize that the first matrix element can be written as
\bea
&& {}_{c_0} F^{\cal O}_{a_1\ldots a_n}(\th|\th_1,\ldots,\th_n)=
F^{\cal O}_{\bar{c}_0a_1\ldots
a_n}(\th+i\pi,\th_1,\ldots,\th_n)+
\lab{ffcross}\\
&& + 2\pi\sum_{i=1}^n\delta_{c_{i-1}a_i}\delta(\th-\th_i)
\left[\prod_{k=1}^{i-1}S_{a_kc_k}^{d_kc_{k-1}}(\th_k-\th)\right]
F^{\cal O}_{d_1\ldots d_{i-1}a_{i+1}\ldots a_n}(\th_1,\ldots\th_{i-1},
\th_{i+1},\ldots,\th_n)\,\,\,.\nonumber
\eea
This result is represented pictorially in fig.A1. As an illustration
of this crossing procedure we write down esplicitely the two matrix
elements involved in the evaluation of the mass and amplitude
variations (\ref{dm}) and (\ref{ds})
\EQ
{}_bF^{\cal O}_a(\th_2|\th_1)=F^{\cal O}_{\bar{b}a}(\th_2+i\pi,\th_1)+
2\pi\delta_{ab}\delta(\th_1-\th_2)\langle0|{\cal O}|0\rangle\,\,,
\EN
\bea
&& {}_{cd}F^{\cal O}_{ab}(\th_3,\th_4|\th_1,\th_2) =
F^{\cal O}_{\bar{c}\bar{d}ab}(\th_3+i\pi,\th_4+i\pi,\th_1,\th_2) + 
\nonumber \\
&& \hspace{15mm}
2\pi\delta(\th_1-\th_3)S_{\bar{d}a}^{ec}(i\pi+\th_4-\th_3)
F^{\cal O}_{eb}(\th_4+i\pi,\th_2) + \nonumber \\
&& \hspace{15mm} + 2\pi\delta(\th_2-\th_4)S_{ab}^{ed}(\th_1-\th_2)
F^{\cal O}_{\bar{c}e}(\th_3+i\pi,\th_1) + \\
&& \hspace{15mm}
+ (2\pi)^2\delta(\th_2-\th_4)\delta(\th_1-\th_3)S_{ab}^{cd}(\th_1-\th_2)
\langle 0|{\cal O}|0\rangle\,\,\,. \nonumber
\eea
In the last equation we excluded the ``resonances'' $\th_4=\th_1$ and
$\th_3=\th_2$.

\appsection

In this section, we collect the needed FF of the operator $\Phi_{1,2}$ 
relative to the $\Phi_{1,3}$ deformation of the minimal 
non-unitarity model $M_{(2,7)}$. They are given by 
\EQ
\langle 0|\Phi_{1,2}|0\rangle=\frac{c_0}{2\cos\frac{2\pi}{5}}\,\,,
\EN
\EQ
F_{11}(\th)= - c_0 4\sin\frac{\pi}{5} 
\frac{F_{11}^{{min}}(\th)}{\P_{\frac{2}{5}}(\th)}\,\,,
\EN
\EQ
F_{22}(\th)= - c_0 8\sin\frac{2\pi}{5} 
\frac
{(\cosh\th + 4 \cos\frac{\pi}{5} \sin^2 \frac{2\pi}{5}) 
F_{22}^{{min}}(\th)}
{\P_{\frac{2}{5}}(\th) \P_{\frac{3}{5}}(\th) \P_{\frac{4}{5}}(\th)}\,\,,
\EN 
and
\EQ
F_{1111}(\th_{ij})=c_0
\frac{\sin^4\frac{\pi}{5}}{\cos^7 \frac{\pi}{5} G^2_{\frac{2}{5}}(0)}
\Q(\th_{ij}) \prod_{l<k}^4 \frac{F_{11}^{{min}}(\th_{lk})}
{\cosh\frac{\th_{lk}}{2} \P_{\frac{2}{5}}(\th_{lk}) }\,\,.
\lab{fourff}
\EN
In the above formulas, the functions 
\EQ
F_{11}^{{min}}(\th)= -i \sinh\frac{\th}{2} G_{\frac{2}{5}}(\th)
\EN
and
\EQ
F_{22}^{{min}}(\th)= -i \sinh\frac{\th}{2} 
G_{\frac{2}{5}}(\th) G_{\frac{3}{5}}(\th) G_{\frac{4}{5}}(\th)\,\,,
\EN
where 
\EQ
G_{\al}(\th)= \exp
\left[2\int_0^{+\infty}\frac{dt}{t} \frac{\cosh(\al-\frac{1}{2})t}
{\cosh\frac{t}{2} \sinh t} \sin^2 \frac{(i\pi-\th)t}{2\pi}\right]\,\,,
\EN
solve the monodromy problem for the form factors, while the functions 
\EQ
\P_{\al}(\th)=\frac{\cos\pi\al -\cosh\th}{2 \cos^2\frac{\pi\al}{2}}
\EN
appearing in the denominators are in coincidence with the bound state 
structure of the theory. The function $\Q$ in the four particle form factor is 
defined as 
\begin{eqnarray*}
\lefteqn{\Q(\th_{ij}) =} \\
& & \left(\sum_{l<k}^4 \cosh\th_{lk} + 4\sin^2 \frac{2\pi}{5}\right)
\left[\cosh\frac{\th_{12}}{2} \cosh\frac{\th_{34}}{2} + 
\cosh\frac{\th_{13}}{2} \cosh\frac{\th_{24}}{2} +
\cosh\frac{\th_{14}}{2} \cosh\frac{\th_{23}}{2}\right] + \\
& & \cos\frac{\pi}{5} \left[\cosh\frac{1}{2}(\th_{12}+\th_{13}+\th_{14}) +
\cosh\frac{1}{2}(\th_{21}+\th_{23}+\th_{24}) + \right.\\
& & \left.\cosh\frac{1}{2}(\th_{31}+\th_{32}+\th_{34}) +
\cosh\frac{1}{2}(\th_{41}+\th_{42}+\th_{43})\right]\,\,.
\end{eqnarray*}
Notice that the constant $c_0$ in front of all Form Factors 
depends on the (non-universal) normalisation\footnote{
The value of $c_0$ corresponding to the standard CFT normalisation, identified 
by the two-point function behaviour  
$<\Phi_{12}(x)\Phi_{12}(0)>\stackrel{x\to 0}{\goto} x^{-4\Delta_{12}}$,
can be hardly determined without resumming the entire spectral series 
of the two-point function. This turns out to be in general quite a 
difficult problem.} of the field $\Phi_{1,2}(x)$. We can get around 
the difficult problem of determining this constant by considering universal 
ratios.

\newpage

{\bf Figure Caption}

\vspace{5mm}

\begin{description}
\item [Figure 1]. Analytic structure of two-body $S$-matrix in non-integrable 
field theories. The crosses indicate the location of the bound state poles 
whereas the thick lines the branch cuts originating at the thresholds.
\item [Figure 2]. Renormalization Group trajectories associated to the 
field theories: integrable directions (continous lines) and 
non-integrable one (dashed line).
\item [Figure 3]. First energy levels of the $\Phi_{1,3}$ integrable 
deformation of the conformal model $M_{(2,7)}$ as functions of the 
strip width $R$. 
\item [Figure 4]. Mass ratio of the non-integrable deformations 
of the conformal model $M_{(2,7)}$ as function of $\chi$.  
\item [Figure 5]. Phase space in the vicinity of the critical point of 
the 2-d Ising model. 
\item [Figure 6]. Analytic structure in momentum space of the 
spin-spin correlation function 
of the Ising model relative to the points ($a,\ldots,g$) in Fig. 5.    
\item [Figure 7]. Numerical spectrum of the Ising model relative to the 
points ($a,\ldots,g$) in Figure 5.  
\item [Figure 8]. (a) The energy level corresponding to a stable
particle above threshold crosses several momentum lines in the
integrable theory; (b) removal of the crossing points implying the
decay of the particle above theshold in the perturbed, non--integrable
theory. 
\item [Figure 9]. Crossing relation for the Form Factors.  
\end{description}

\clearpage
\pagestyle{empty}

\vfill
\begin{figure}
\centerline{
\epsfxsize=400pt
\epsfbox{figure12.eps}}
\end{figure}\vfill


\clearpage\vfill
\begin{figure}
\centerline{
\epsfxsize=400pt
\epsfbox{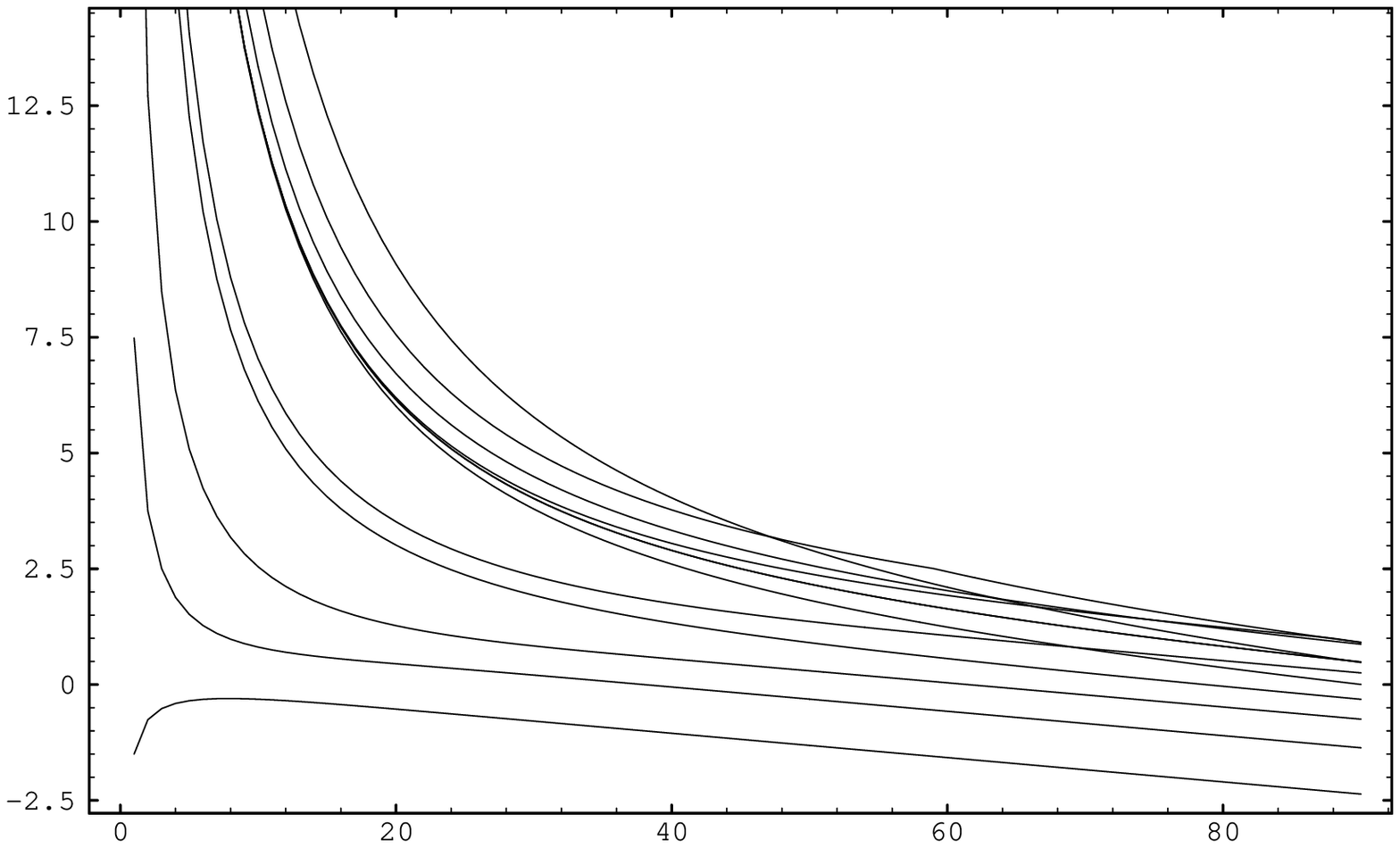}}
\centerline{{\bf Figure 3}}
\end{figure}\vfill

\clearpage\vfill
\begin{figure}
\centerline{
\epsfxsize=400pt
\epsfbox{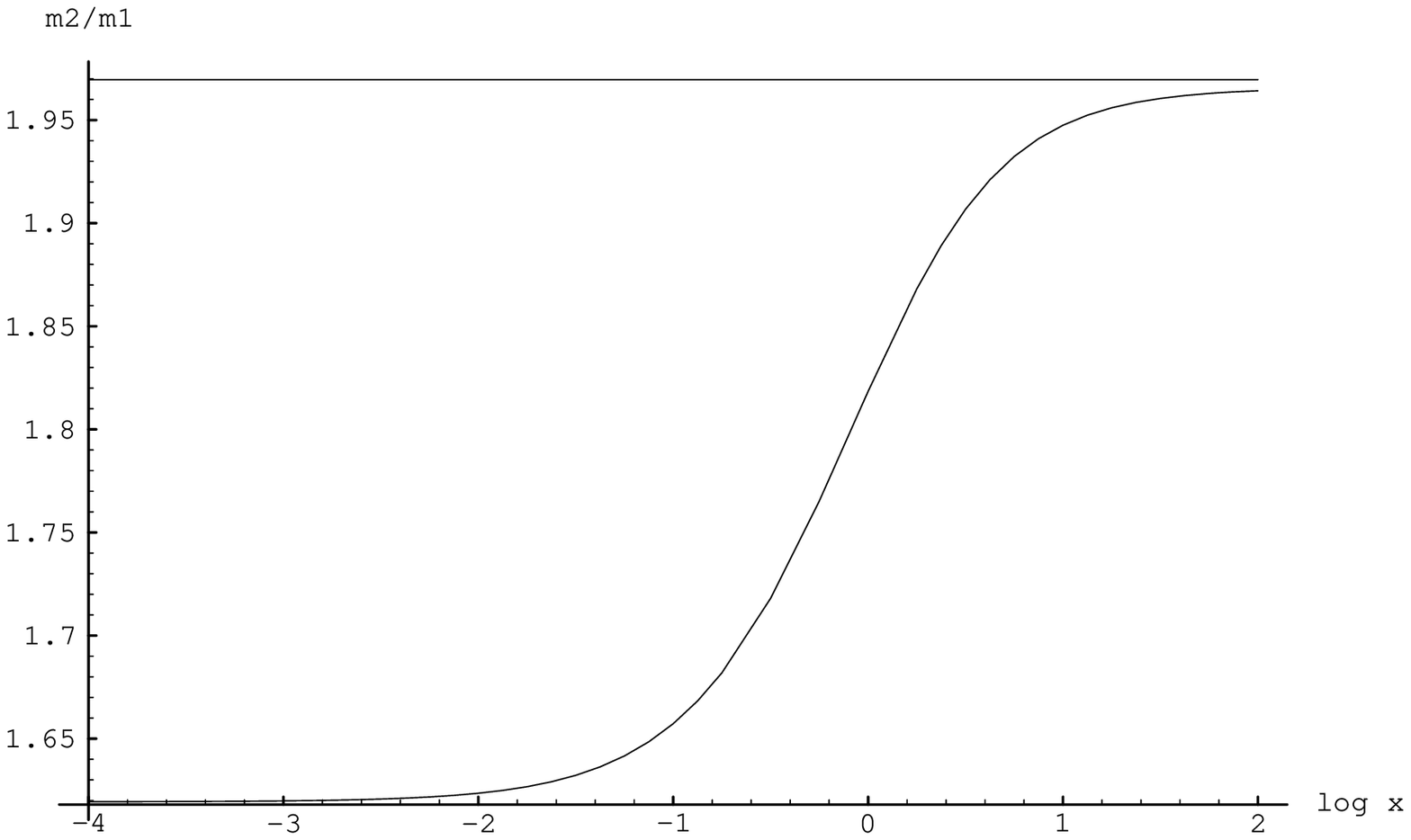}}
\centerline{{\bf Figure 4}}
\end{figure}\vfill

\clearpage\vfill
\begin{figure}
\centerline{
\epsfxsize=500pt
\epsfbox{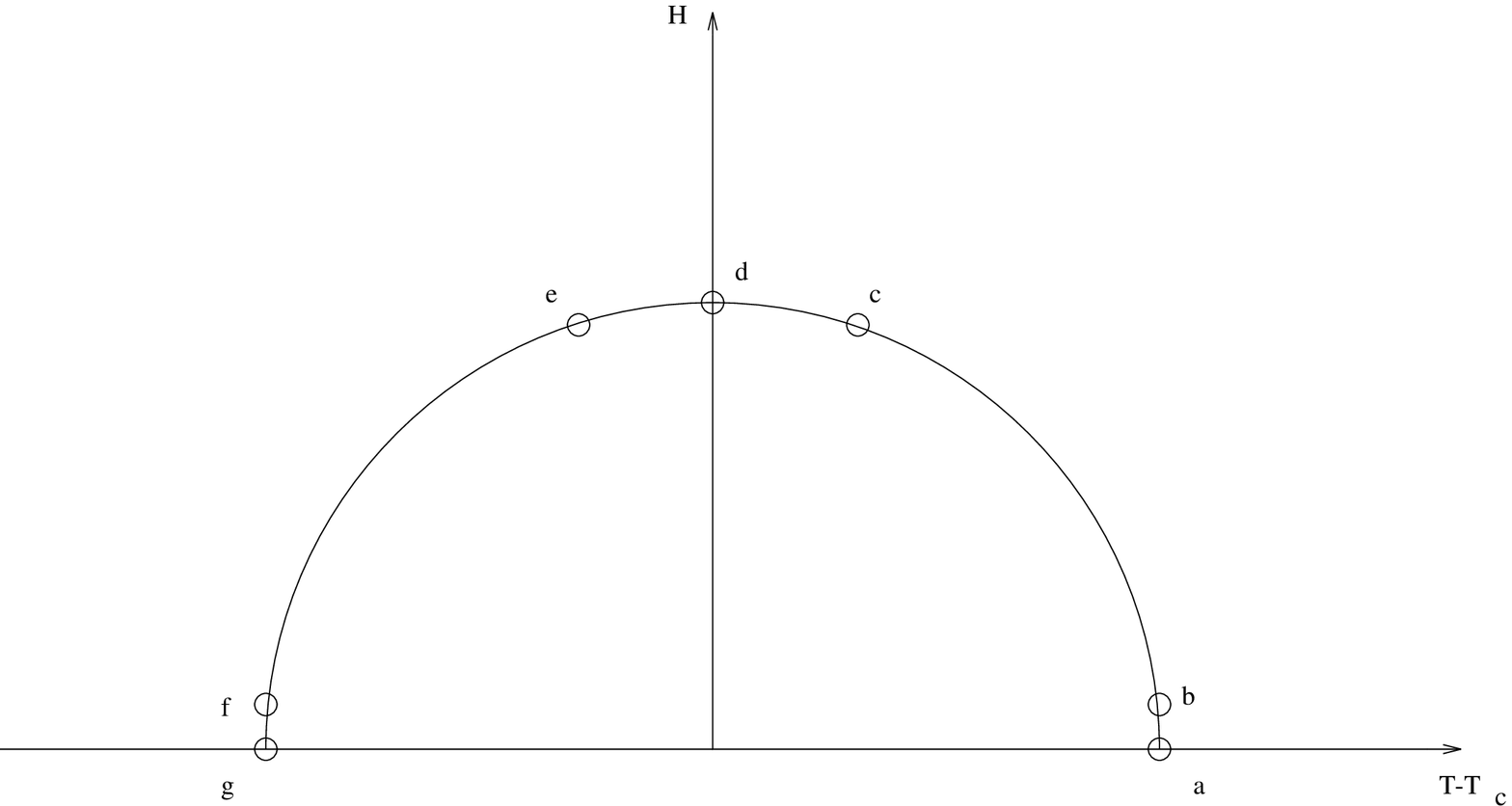}}
\centerline{{\bf Figure 5}}
\end{figure}\vfill

\clearpage\vfill
\begin{figure}
\centerline{
\epsfxsize=300pt
\epsfbox{figure6a.eps}}
\end{figure}\vfill

\clearpage\vfill
\begin{figure}
\centerline{
\epsfxsize=300pt
\epsfbox{figure6b.eps}}
\end{figure}\vfill

\clearpage\vfill
\begin{figure}
\centerline{
\epsfxsize=400pt
\epsfbox{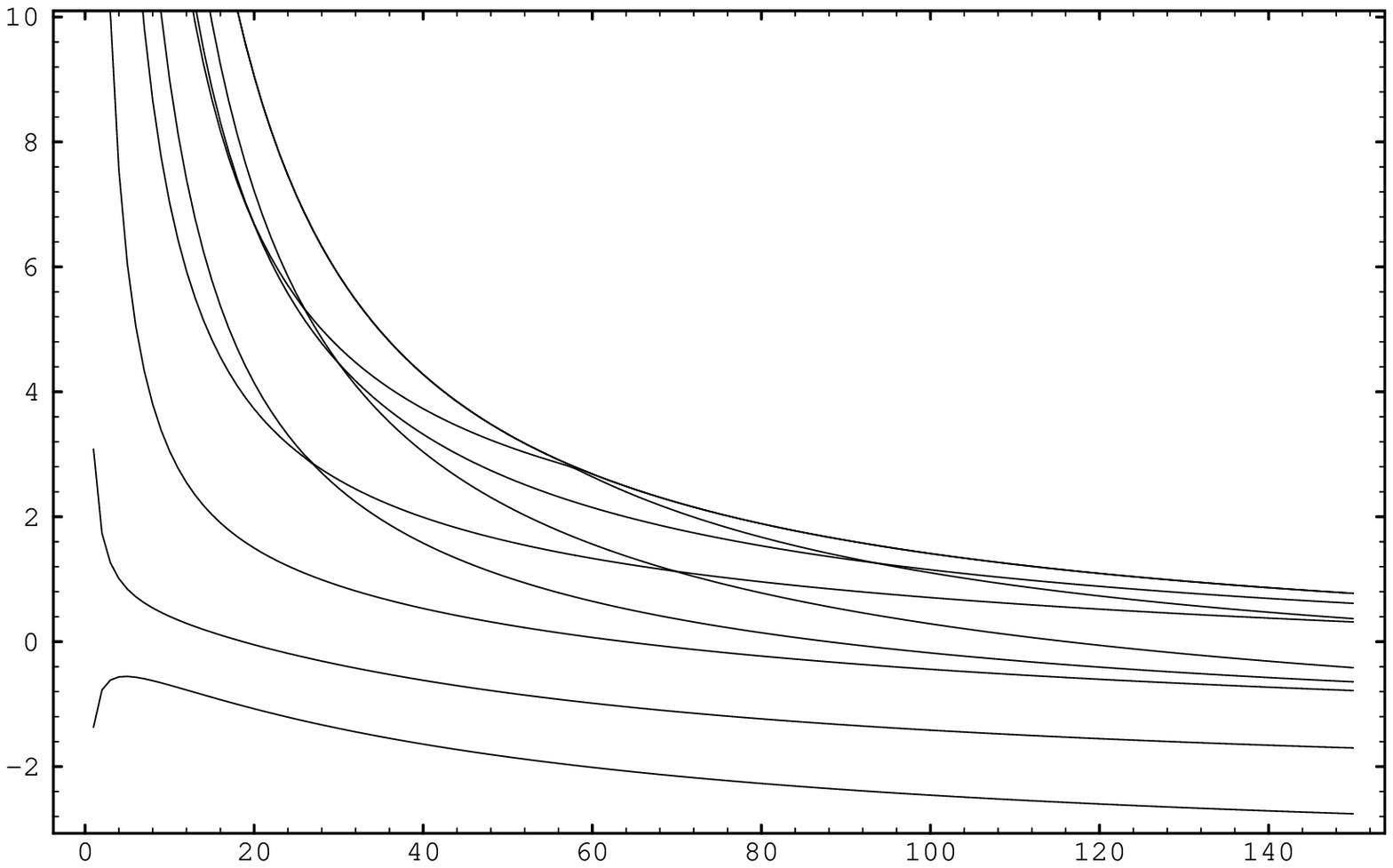}}
\centerline{{\bf Figure 7.a}}
\end{figure}\vfill

\clearpage\vfill
\begin{figure}
\centerline{
\epsfxsize=400pt
\epsfbox{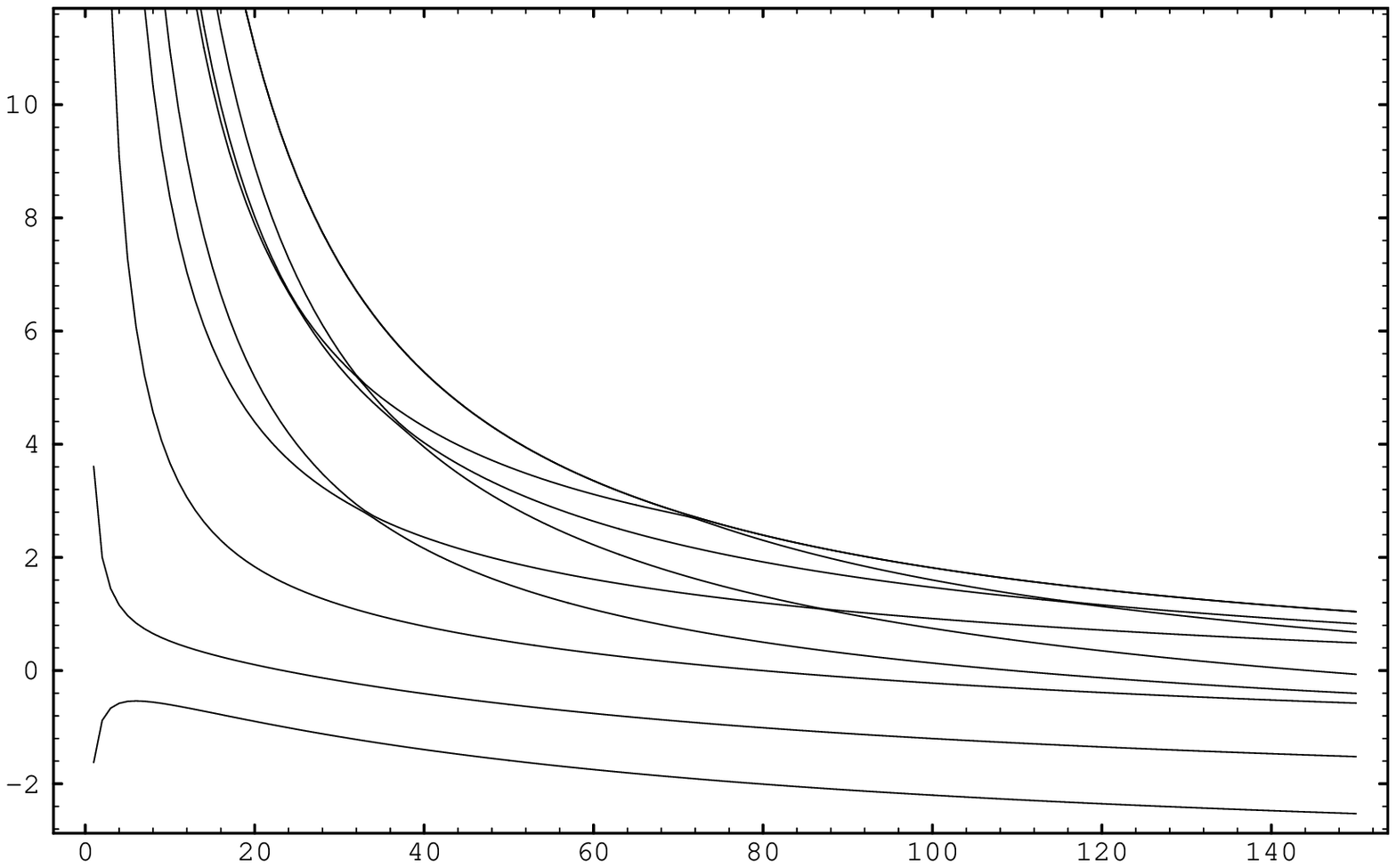}}
\centerline{{\bf Figure 7.b}}
\end{figure}\vfill

\clearpage\vfill
\begin{figure}
\centerline{
\epsfxsize=400pt
\epsfbox{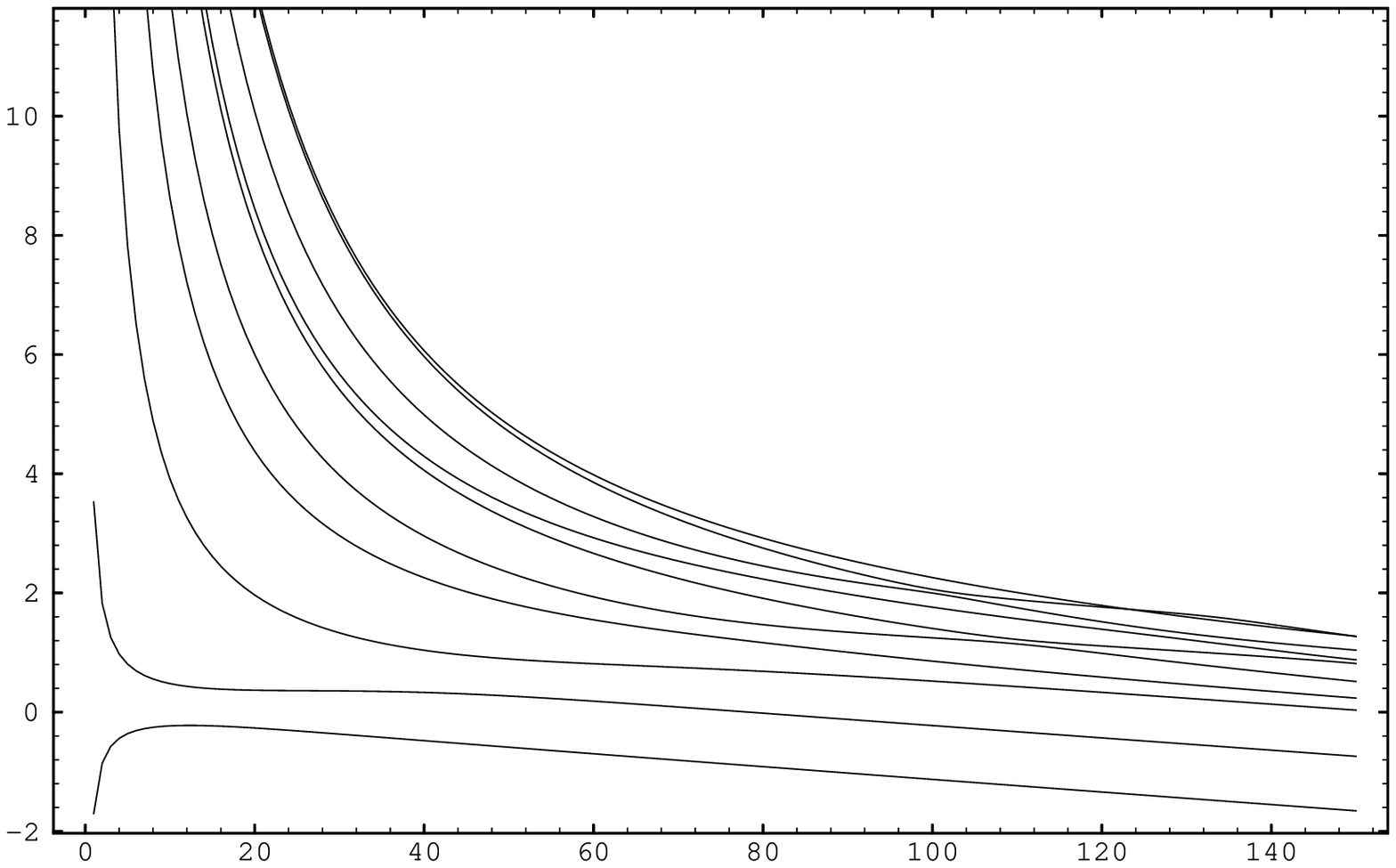}}
\centerline{{\bf Figure 7.c}}
\end{figure}\vfill

\clearpage\vfill
\begin{figure}
\centerline{
\epsfxsize=400pt
\epsfbox{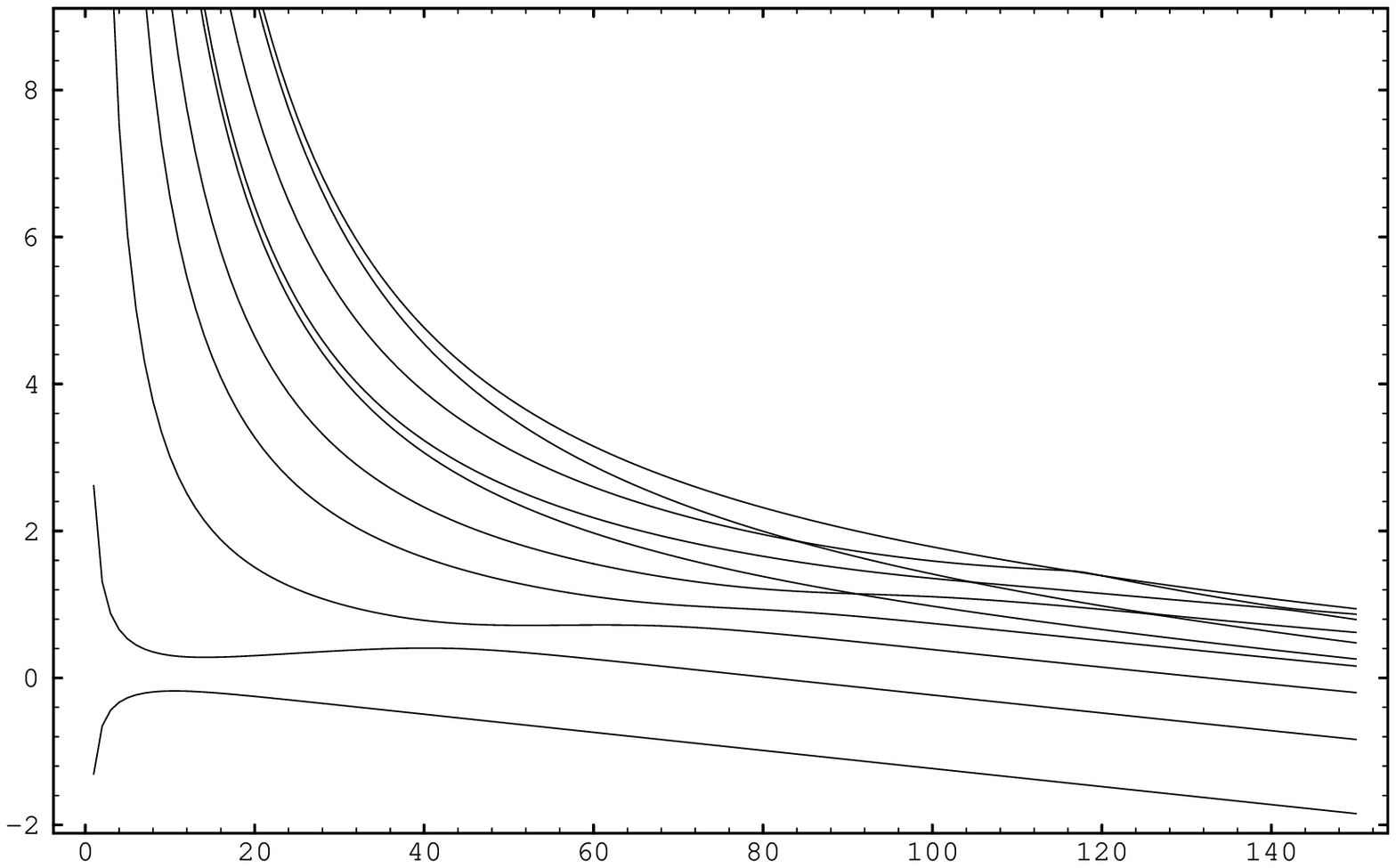}}
\centerline{{\bf Figure 7.d}}
\end{figure}\vfill

\clearpage\vfill
\begin{figure}
\centerline{
\epsfxsize=400pt
\epsfbox{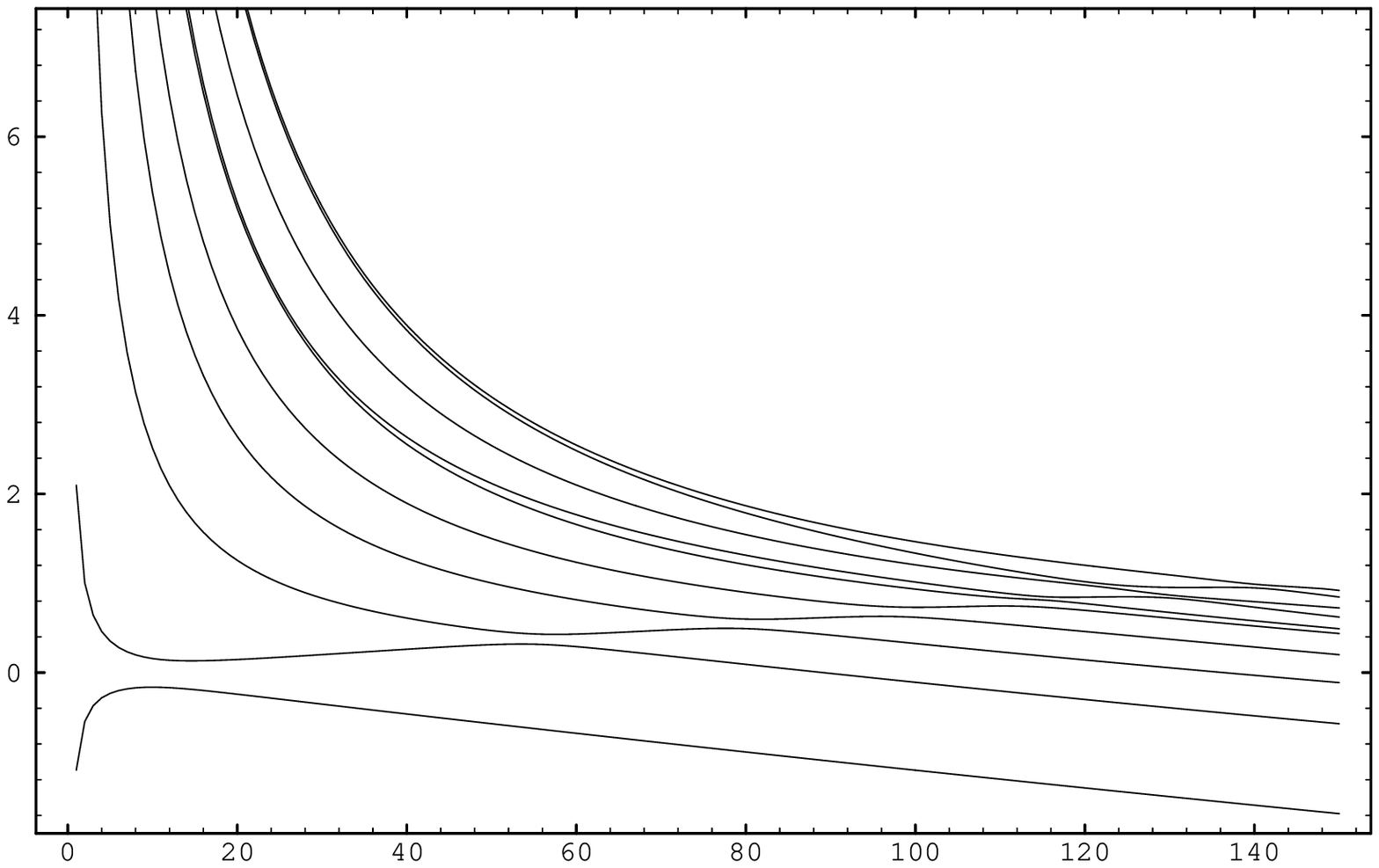}}
\centerline{{\bf Figure 7.e}}
\end{figure}\vfill

\clearpage\vfill
\begin{figure}
\centerline{
\epsfxsize=400pt
\epsfbox{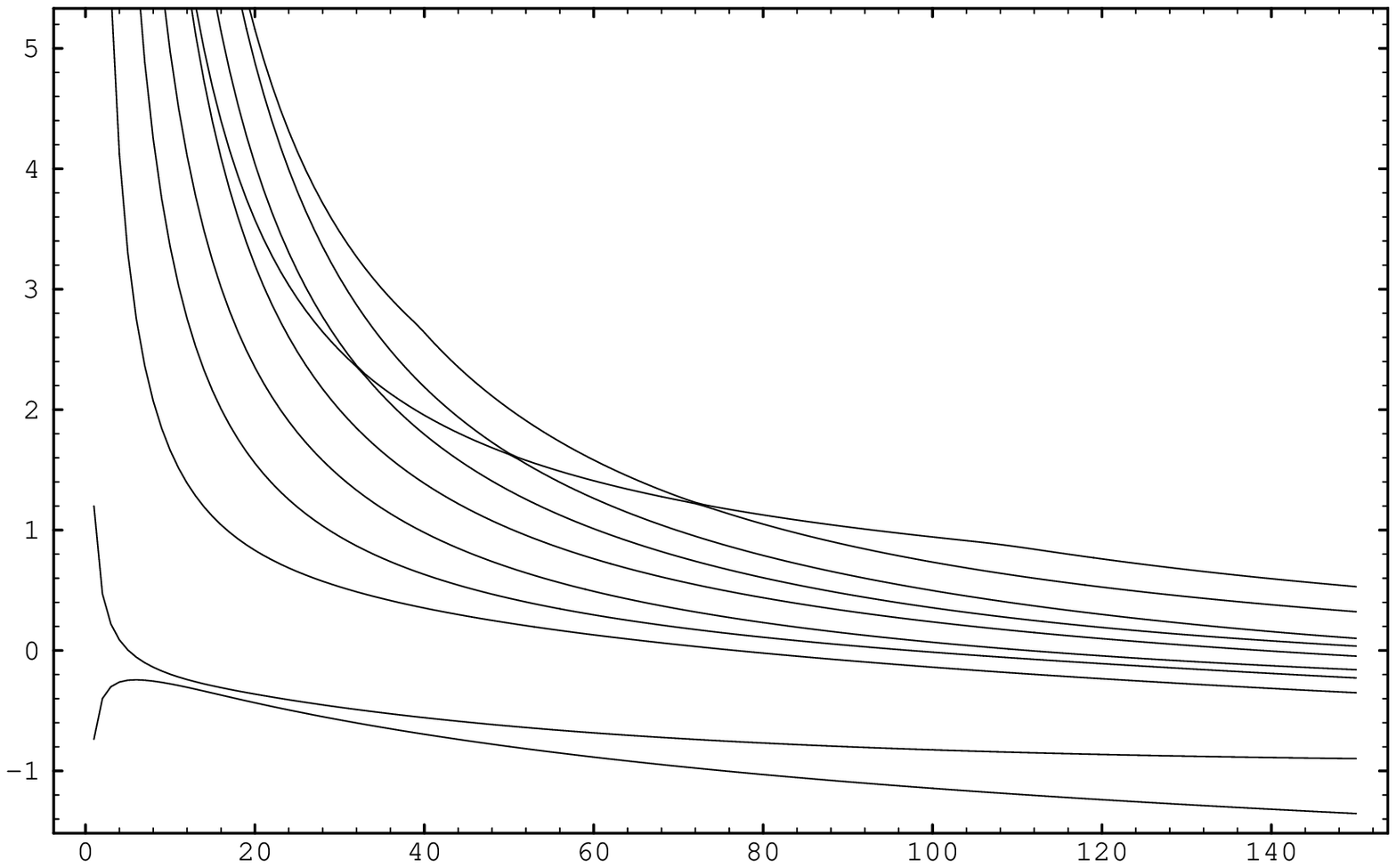}}
\centerline{{\bf Figure 7.f}}
\end{figure}\vfill

\clearpage\vfill
\begin{figure}
\centerline{
\epsfxsize=400pt
\epsfbox{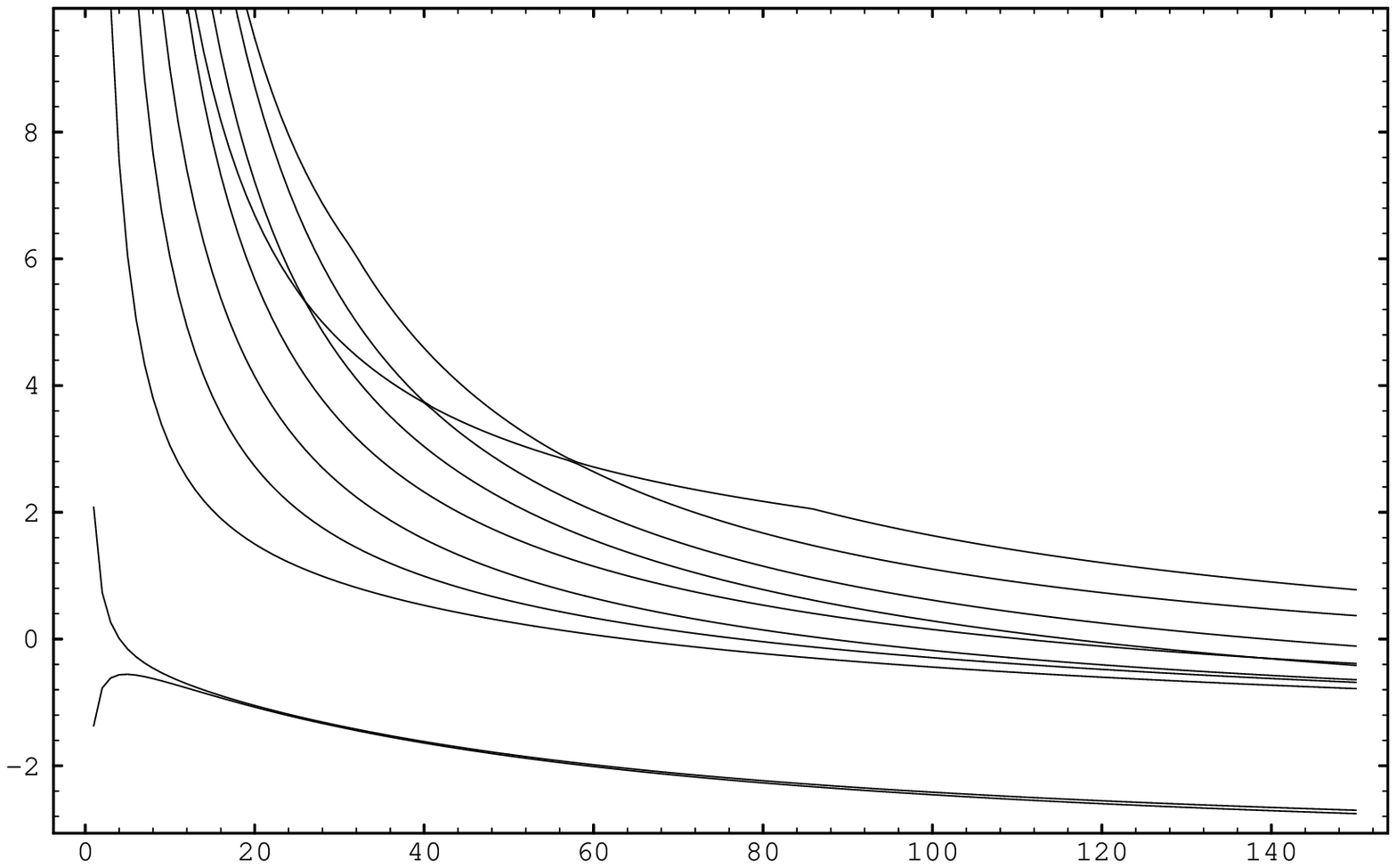}}
\centerline{{\bf Figure 7.g}}
\end{figure}\vfill

\clearpage\vfill
\begin{figure}
\centerline{
\epsfxsize=400pt
\epsfbox{figure8.eps}}
\centerline{{\bf Figure 8}}
\end{figure}\vfill

\clearpage\vfill
\begin{figure}
\centerline{
\epsfxsize=500pt
\epsfbox{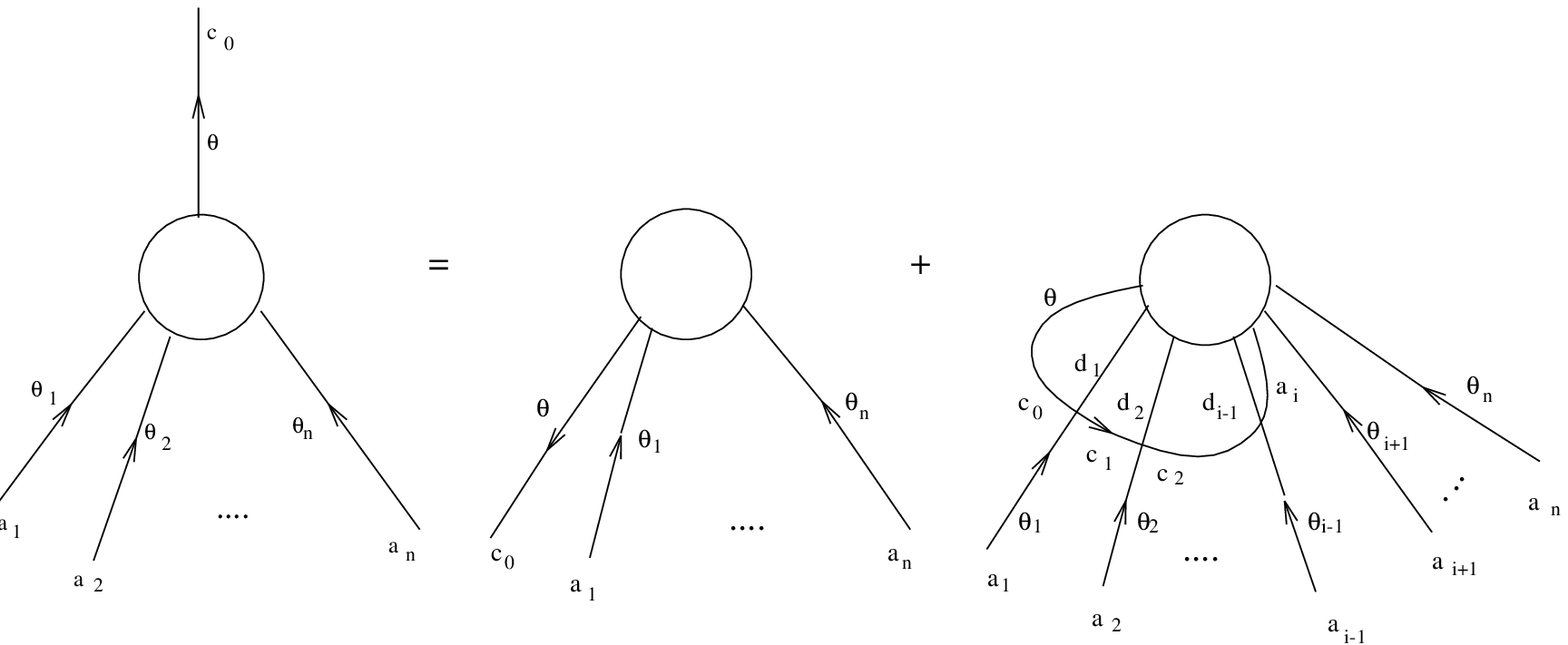}}
\centerline{{\bf Figure 9}}
\end{figure}\vfill

\end{document}